\documentclass[onecolumn]{mn2e}
\pdfoutput=1
\usepackage{graphicx,amsmath}
\usepackage{txfonts}
\usepackage{booktabs}

\title[Imprints of Dark Energy on Cosmic Structure Formation
I) Realistic Quintessence Models]{Imprints of Dark Energy on Cosmic Structure Formation \\ I) Realistic Quintessence Models and the Non-Linear Matter Power Spectrum}  
\author[J.-M. Alimi, A. F\"uzfa, V. Boucher, Y. Rasera, J. Courtin, P.-S. Corasaniti]{
J.-M. Alimi$^{1,2}$\thanks{email: jean-michel.alimi@obspm.fr}, A. F\"uzfa$^{1,2,3}$\thanks{email: andre.fuzfa@fundp.ac.be}, V. Boucher$^{3}$\thanks{email: vincent.boucher@uclouvain.be}, Y. Rasera$^{1}$\thanks{email: yann.rasera@obspm.fr}, J. Courtin$^{1}$\thanks{email: jerome.courtin@obspm.fr}, P.-S. Corasaniti$^{1}$\thanks{email: pier-stefano.corasaniti@obspm.fr}\\
$^{1}$CNRS, Laboratoire Univers et Th\'eories (LUTh), UMR 8102 CNRS, Observatoire de Paris,\\
\quad Universit\'e Paris Diderot ; 5 Place Jules Janssen, F-92190 Meudon, France\\
$^{2}$Groupe d'Application des MAth\'ematiques aux Sciences du COsmos (GAMASCO), \\
\quad University of Namur (FUNDP), 61 rue de Bruxelles, B-5000 Namur, Belgium\\
$^{3}$ Center for Particle Physics and Phenomenology (CP3),\\
\quad Universit\'e catholique de Louvain, 2 Chemin du Cyclotron, B-1348 Louvain-la-Neuve, Belgium} 
    
\begin{document}  
  
%\date{Accepted ... Received ... ; in original form ...}  
  
\pagerange{}
 \pubyear{2009}  
  
\maketitle  
  
\label{firstpage}  
  
\begin{abstract}  
Quintessence has been proposed to account for dark energy 
in the Universe. This component causes a typical modification of the
background cosmic expansion, which in addition to its clustering
properties, can leave a potentially distinctive signature on
large scale structures. Many previous studies have investigated this
topic, particularly in relation to the non-linear regime of
structure formation. However, no careful pre-selection of viable 
quintessence models with high precision cosmological data
was performed. Here we show that this has led to a 
misinterpretation (and underestimation) of the imprint of 
quintessence on the distribution of large scale structures.
To this purpose we perform a likelihood analysis of the 
combined Supernova Ia UNION dataset and WMAP5-years data to
identify realistic quintessence models. These are specified
by different model parameter values, but still
statistically indistinguishable from the vanilla $\Lambda$CDM.
Differences are especially manifest
in the predicted amplitude and shape of the linear matter power
spectrum, though these remain within the uncertainties 
of the SDSS data. We use these models as benchmark for studying the clustering 
properties of dark matter halos by performing a series of high resolution
N-body simulations. In this first paper,
we specifically focus on the non-linear matter power spectrum. 
We find that realistic quintessence models allow for relevant
differences of the dark matter distribution with the respect to the
$\Lambda$CDM scenario well into the non-linear regime, with deviations 
up to $40\%$ in the non-linear power spectrum. 
Such differences are shown to depend on the nature of DE, as well as the
scale and epoch considered. At small scales 
($k\sim 1-5$ h $\rm Mpc^{-1}$, depending on the redshift) 
the structure formation process is about $20\%$ more efficient 
than in $\Lambda$CDM. We show that these imprints are a specific record of the
cosmic structure formation history in dark energy cosmologies and
therefore cannot be accounted in standard fitting functions of the 
non-linear matter power spectrum.
\end{abstract}  


\newpage
\section{Introduction}
Over the past decade cosmological observations have provided mounting
evidence in favour of an unexpected energy component --- dark energy (DE)
--- which dominates the present energy content of the Universe and is
responsible for the recent cosmic accelerated expansion. Early measurements
of the Supernova Ia Hubble diagram (Riess et al. 1998, 2001; Perlmutter et al. 1999)
which have been recently 
confirmed by more accurate detections (Knop et al. 2003; Astier et al.
2006; Riess et al. 2007), in combination with precise measurements of Cosmic
Microwave Background (CMB) anisotropies (De Bernardis et al. 2000; Spergel et al. 2003, 2006; Bennett et al.
2003; Komatsu et al. 2008), mapping of the distribution of large scale
structures from galaxy survey (Efsthathiou et al. 2002; Tegmark et al. 2004, 2006; 
Cole et al. 2005; Eisenstein et al. 2005), CMB--Large Scale Structure (LSS) correlation (see Cabre et al. 2006, Giannantonio et al.
2008, McEwen et al. 2008 and references therein), galaxy clusters (Allen, Schmidt \& Fabian 2002), 
peculiar velocities (Mohayaee \& Tully 2005) and galaxy redshift distortions 
(Guzzo et al. 2008) have pinned down the cosmic abundance of the 
various matter energy components. 
These data indicate that pressureless matter (baryons and cold dark matter)
accounts for only $~25\%$ of the total energy content in the Universe, 
leaving the
remaining $~75\%$ to be dark energy. 

In light of these observations the existence of DE phenomenon can 
hardly be contested nowadays, thus the question of its physical origin has become crucial especially 
from a fundamental physics perspective. The simplest scenario to account for a late time cosmic 
acceleration consists of a positive cosmological constant $\Lambda$ in Einstein's equations of
general relativity. Nevertheless, the fact that today $\rho_\Lambda$ is of the order of the
matter density $\rho_m$ suggests that we are living in a period of remarkable coincidence,
since the value of $\Lambda$ must have been fixed very precisely at
early times as to allow for a sufficient period of structure formation
responsible for the distribution of matter that we observe today
(see e.g. Weinberg 1989). One can hope that understanding the physical 
nature of $\Lambda$ may solve this puzzling coincidence, 
but all known attempts have led to an even more troubling problem 
that can be explained only if an anthropic selection mechanism is at work. 
In fact the cosmological constant can be naturally interpreted as the energy
contribution of quantum vacuum fluctuations, however Standard Model
fields give rise to a vacuum energy density up 
to 119 order of magnitudes (assuming a cut-off at the Planck scale) 
larger than the cosmological constant observed value, hence requiring
an unnatural fine-tuning of the bare cosmological constant such as to ensure a precise cancelation
of the energy associated to quantum vacuum diagrams. We still lack of a convincing theoretical explanation
for this naturalness problem, and it may well be that the dark energy
phenomenon is of complete different origin. 

Several scenarios have been proposed
in a vast literature, which include the existence of a light minimally coupled scalar field
dubbed {\it quintessence} (Wetterich 1988; Ratra \& Peebles 1988) 
motivated by physics beyond the Standard Model
of particle physics, deviations from standard general relativity on cosmological scales due to
higher order corrections to Einstein gravity (see e.g. Capozziello, Cardone \& Troisi 2005), 
and modifications due to the presence of extra dimensions (Dvali, Gabadadze and Porrati 2000).
In quintessence cosmologies the late time dynamics of the scalar field is responsible for
driving the cosmic accelerated expansion (see Copeland, Sami \& Tsujikawa 2006 for a review). 
As the field rolls down its self-interaction potential
the kinetic energy becomes small compared to the potential energy, causing the field
pressure becoming sufficiently negative such as to drive the acceleration, namely 
$w_{DE}=p_{DE}/\rho_{DE}<-1/ (3\Omega_{DE})$ (for a flat universe) at present time, 
where $\Omega_{DE}=8\pi \rho_{DE}/(3m_{Pl}^2H_0^2)$ is the density parameter (with $m_{Pl}$ 
being the Planck mass and $H_0$ the Hubble constant). A particular class of quintessence models
is characterized by ``{\it tracking}'' potentials, for which the field dynamics has 
a late time accelerating attractor solution, preceeded by a period during which 
the scalar field energy density tracks that of the dominant background component 
(see Steinhardt, Wang \& Zlatev 1999; Zlatev, Wang \& Steinhardt 1999).
In such models any dependency on the initial conditions is therefore erased, thus
alleviating the coincidence problem. Along these lines a full solution 
to such coincidence is provided by non-minimally coupled scalar field models,
where the matter components are directly coupled to the scalar field (see e.g.
Amendola 2000; Khoury \& Weltman 2004; Alimi \& F\"uzfa 2008) 

Precision cosmology offers a unique opportunity to test the nature of dark energy.
However, the parameter inference is usually limited by degeneracies amongst the 
various cosmological parameters, and the combination of several probes is
indeed necessary. The imprint of quintessence on the CMB anisotropy power spectrum
and the matter power spectrum has been studied in several works 
(e.g. Viana \& Liddle 1998; Caldwell et al. 1998; Perrotta \& Baccigalupi 1999;
Brax, Martin \& Riazuelo 2000). In particular it was shown that a dynamical dark energy component
leaves a distinct signature on the CMB through the Integrated Sachs-Wolfe (ISW) effect (Corasaniti et al. 2003),
such that models fitting the CMB data, can allow for very different values of the normalization
$\sigma_8$ of the linear matter power spectrum (Kunz et al. 2004). The viability of these models has
been tested against CMB and SN Ia data in various works (Corasaniti \& Copeland 2002; 
Baccigalupi et al. 2002; Colombo \& Gervasi, 2006). These models can provide fit to 
the data at the same statistical level of the $\Lambda$CDM scenario, implying 
that cosmological data at redshift higher than $z\approx 1$ are
therefore necessary to further break parameter degeneracies and
possibly distinguish between competing DE models.

In recent years there has been a growing interest on
whether the clustering of large-scale structures might be
crucial for settling the debate on the physical nature of DE.
As this process covers a long-period between the dark ages and today
($0<z<1000$), it appears as a unique experimentation field for building
new tests of DE. Several studies have performed N-body simulations to
evaluate the imprint of DE on the non-linear regime of gravitational 
collapse and determined the signature on the mass power spectrum 
and cluster mass function (see e.g. Ma et al. 1999; Bode et al. 2001; 
Lokas, Bode \& Hoffman 2004; Munshi, Porciani \& Wang 2004). 
These works have focus on DE models specified by a constant equation 
of state, while Benabed \& Bernardeau (2001) have performed an analytical
study of the growth of matter pertubations for quintessence models
at linear and second order respectively, and estimated the effect
on the non-linear power spectrum. N-body simulations dedicated to quintessence 
cosmologies have been performed in various works 
(Klypin et. 2003; Dolag et al. 2004; Solevi et al. 2006),
other studies have also included the gas 
through N-body hydrodynamical simulations (Maio et al. 2006;
Aghanim, da Silva \& Nunes 2008). These studies have concluded that 
the imprint of dark energy on the non-linear matter power spectrum
and the cluster mass function at $z=0$ remains indistinguishable
from the $\Lambda\rm CDM$, while significant differences 
may arise at higher redshifts, with different halo profiles at all redshifts.

A feature common to all these works is the fact that these analysis have 
exclusively focused on models specified by the same cosmological parameters values,
independently of whether these values provide compatible fits of the underlying
dark energy model to existing data\footnote{A noticeable exception is the work of Casarini et al. 2009
in which one of the models considered is compatible with recent data, as derived in La Vacca et al. 2009. 
However, their approach is rather different than ours
and we will discuss some of their results in section 5.5.}. This is a crucial point, since it does not allow us
to address an important question, namely whether detailed measurements of the non-linear
clustering of matter are sensitive to the imprint quintessence/dark energy 
models which are currently not distinguishable from the standard $\Lambda \rm CDM$
using SN Ia, CMB and linear matter power spectrum data. This is particularly relevant
given the upcoming observational campaigns which will probe with high precision
several aspects of the matter clustering. This question will
be addressed in this paper, the first of a series
dedicated to the imprint of ``realistic'' dark energy models on the non-linear
structure formation. In the upcoming papers we will mostly focus on the physical properties
of DM halos: mass functions, density profiles, virialisation process and internal
velocity distribution. We will also investigate the spatial halo
distribution and the influence on cosmological bias. 

The paper is organized as follows: in Section~\ref{quinte_eq} we recall
the basic equations describing the dynamics of quintessential
cosmologies at the background and linear perturbation level;
in Section~\ref{like_res} we present the results of a combined
likelihood analysis of the WMAP5-years data and SN Ia from
the UNION dataset to constrain two different 
types of quintessence models specified by Ratra-Peebles and SUGRA 
potentials respectively. We then identify \textit{realistic} quintessence models
which are statistically indistinguishable from the vanilla $\Lambda \rm CDM$ and discuss
their main phenomenological features in Section~\ref{quinte_impr}. 
We use these realistic models as benchmark for studying the structure
formation through state-of-the-art N-body simulations
which we present in Section~\ref{quinte_sim}. We will describe the numerical software
developed for the realization of the project, the characteristics of the simulations
and discuss the results. In particular we find that 
quintessence leaves several imprints in the non-linear matter power spectrum 
at all redshifts and on the non-linear growth of cosmic structures.
Finally we will present our conclusion and discuss future work
in Section~\ref{conclu}.

%__________________________________________________________________

\section{Quintessential Cosmology}\label{quinte_eq}
\subsection{Basic Homogeneous Equations}
Assuming the cosmological principle, the large scale Universe can be
considered homogeneous and isotropic and can thus be 
described by the Friedmann-Lema\^{i}tre-Robertson-Walker (FLRW) metric, 
which in terms of the conformal time $\eta$ reads as:
\begin{equation}
\label{flrw}
ds^2=a^2(\eta)\left[-d\eta^2+dl^2\right],
\end{equation}
where $a(\eta)$ is the scale factor and $dl$ is the length element. In
this paper, we restrict to spatially flat geometries. Throughout
the paper we will use Planck units ($\hbar=c=1$, $m_{Pl}=1/\sqrt{G}=1.2211\times 10^{19} GeV$).
We assume the cosmic matter content to consist of pressureless matter (composed of ordinary 
baryonic matter and cold dark matter), relativistic matter (photons
and neutrinos) and a quintessence scalar field. 
The expansion rate of the Universe, defined in terms of the scale
factor by the \textit{Hubble parameter} $\mathcal{H}=a'/a$ 
(a prime denotes a derivative with respect to the conformal time $\eta$), 
is given by the Friedmann equation:
\begin{equation}
\label{friedmann}
\mathcal{H}^2=H_0^2\left[\frac{\Omega_{m}}{a}+\frac{\Omega_{r}}{a^2}\right]+\frac{8\pi}{3m_{Pl}^2}\rho_Q a^2,
\end{equation}
with $H_0$ the Hubble constant and $\Omega_i=\rho_i/\rho_c$ is the
present value of the density parameters,
where $\rho_c=3H_0^2m_{Pl}^2/(8\pi)$ is the critical density and
$\rho_i=\rho_m,\rho_r$ indicates respectively the 
density contribution of pressureless matter and radiation; $\rho_Q$ 
stands for the energy density of quintessence. 
\\
\\
The quintessence field is assumed to be a neutral (real) scalar field
$\varphi(\eta)$ with self-interaction potential $V(\varphi)$, which couples to ordinary matter
only through its gravitational influence, (i.e. minimally coupled
scalar field). The field dynamics is given by the
\textit{Klein-Gordon} equation:
\begin{equation}
\label{kg}
\varphi''+2\mathcal{H}\varphi'+\frac{a^2}{m_{Pl}^2}\frac{dV(\varphi)}{d\varphi}=0,
\end{equation}
where $\varphi$ is the field expressed in units of the Planck mass. 
The quintessence energy density and pressure reads as:
\begin{align}
\rho_Q&= \frac{m_{Pl}^2}{2a^2}\varphi^{'2}+V(\varphi),\nonumber\\
p_Q&= w_Q \rho_Q=\frac{m_{Pl}^2}{2a^2}\varphi^{'2}-V(\varphi)\cdot\nonumber
\end{align}
where $w_Q$ is the quintessence equation of state.\\
\\
The evolution of the system is completely determined by specifying the
form of the quintessence potential and the scalar field initial
conditions $(\varphi_i,\; \varphi'_i)$, which can be set in the early
Universe, for example at the end of inflation. 
Convenient choices of $V(\varphi)$ are the so-called 
\textit{tracking potentials} (see Steinhardt, Wang \& Zlatev 1999 and references
therein) characterized by a late time scalar field dominated regime,
which can be reached from a large range of initial conditions, spreading over dozens of
order of magnitudes in the field phase space. 

We consider two well-known examples of such potentials. The
\textit{Ratra-Peebles} (RP) inverse power law (Wetterich 1988; Ratra
\& Peebles 1988),
\begin{equation}
\label{rp}
V_{RP}(\varphi)=\frac{\lambda^{4+\alpha}}{m_{Pl}^\alpha\varphi^\alpha},
\end{equation}
where $\alpha\ge 0$ and $\lambda>0$ are free parameters characterizing
the slope and amplitude of the scalar self-interaction. 
Here we fix the energy scale $\lambda$ such that for a given value of
$\alpha$ the input value of $\Omega_Q$ is retrieved. This model was
originally proposed to mimic a time-varying cosmological constant
(Wetterich 1988), which for nearly flat slopes is recovered at late
times, when the scalar field remains frozen at large values by the
cosmic expansion. A quintessence inverse power law potential has also
been motivated in the framework of supersymmetric extensions of QCD
(Binetruy 1999; Masiero, Pietroni and Rosati 2000). However, as
noticed by Brax \& Martin (2000) for present field values
supergravity corrections should be included leading to the following form:
\begin{equation}
\label{sugra}
V_{SU}(\varphi)=\frac{\lambda^{4+\alpha}}{m_{Pl}^\alpha\varphi^\alpha}e^{4\pi\varphi^2},
\end{equation}
which we will refer to as the \textit{SUGRA} model. 

\subsection{Perturbed Quintessential Cosmology}
At the inhomogeneous level, assuming small perturbations about the
FLRW metric, the linearly perturbed line element in the Newtonian
gauge is given by
\begin{equation}
ds^2=a^2(\eta)\left[-(1+2\Phi(\vec x,\eta))d\eta^2+(1-2\Psi(\vec x,\eta))dl^2\right]\cdot
\end{equation}
where $\Phi$ and $\Psi$ are the Bardeen potentials. Differently from
the cosmological constant a quintessential dark energy can cluster and
its density perturbations can contribute to the evolution of the
gravitational potentials in addition to the perturbations in the other
matter components. In order to correctly compute the CMB and the
linear matter power spectra, we have modified the \textsc{CAMB} code 
(Lewis, Challinor \& Lasenby 1999) to account both for the
modification of the background expansion and the clustering of
quintessence. Two equivalent approaches have been adopted to
cross-check the validity of the computation.
The first is to evolve the perturbations $\delta\varphi$ of the 
quintessence field itself through the perturbed Klein-Gordon
equation. 
In the Fourier space we have:
\begin{equation}
 \delta\varphi'' + 2\mathcal{H}\delta\varphi' + \left(k^2+\frac{a^2}{m_{Pl}^2}\dfrac{d^2V}{d\varphi^2}\right)\delta\varphi + 2 \Phi \frac{a^2}{m_{Pl}^2}\dfrac{dV}{d\varphi} - (3\Psi'+\Phi') \varphi' = 0\cdot
 \label{pert_kg}
\end{equation} 
The perturbed Einstein equations give the evolution of gravitational
potentials $\Psi$ and $\Phi$,
\begin{align}
 k^2\Psi + 3\mathcal{H}(\Psi'+\mathcal{H}\Phi) &= - \dfrac{4\pi}{m_{Pl}^2} a^2 \rho_s \delta_s\\
\Psi' + \mathcal{H}\Phi &= - \dfrac{4\pi}{m_{Pl}^2} a^2 \rho_s (1+w_s) v_s\;,
\end{align}
where $\delta_s$, $v_s$ and $w_s$ are the relative density contrast
$\delta\rho_s/\rho_s$, velocity and equation of state for a given
fluid $s$ (see Eq.~(\ref{defs}) for the definition of the quintessence
density contrast and velocity in terms of the scalar field
perturbation). An implicit summation over all the species $s$
(photons, baryons, cold dark matter, neutrinos and quintessence) has been assumed.
Written in terms of the scalar field perturbation the quintessence
density contrast and velocity reads as:
\begin{align}
\frac{a^2}{m_{Pl}^2}\rho_Q\delta_Q&= \varphi'\delta\varphi'+\frac{a^2}{m_{Pl}^2} \frac{dV(\varphi)}{d\varphi}\delta\varphi-\varphi'^2 \Phi\nonumber\\
v_Q&= -\dfrac{\delta\varphi}{\varphi'}\;.
 \label{defs}
\end{align}
The second approach consists in considering quintessence as an
additional fluid, in such a case the quintessence perturbations are
characterized by an adiabatic sound speed
$c_A^2\equiv\frac{p_Q'}{\rho_Q'}$ and an intrinsinc entropy
perturbation $\Gamma_Q\equiv\frac{\delta
  p_Q}{p_Q}-\frac{c_A^2}{w_Q}\delta_Q$. 
In the framework of quintessence models these quantities take the following form
\begin{align}
c_A^2 & =  1+\frac{2}{3m_{Pl}^2}\frac{dV}{d\varphi}\frac{a^2}{\varphi'\mathcal{H}}\;, \\
\Gamma_Q & =  \frac{1-c_A^2}{w_Q}\left[\delta_Q-3\mathcal{H}(1+w_Q)v_Q\right]\;.
\end{align}
Thus, the continuity and the Euler equation simplify to
\begin{align}
\delta_Q' & =  -3\mathcal{H}(1-w_Q)\delta_Q+(1+w_Q)\left[\left(9\mathcal{H}(1-c_A^2)+ k^2\right)v_Q+3\Psi'\right]\nonumber\\
v_Q'& =  -\dfrac{\delta_Q}{1+w_Q}+2\mathcal{H}v_Q -\Phi\cdot \label{pertq1}
\end{align}

For super-Hubble scales, $k\mathcal{H}^{-1}\ll 1$, the tracking
properties of the self-interaction potential force the quintessence
perturbations to follow a tracking solution as well (see Brax, Martin \&
Riazuelo 2000). A departure from this regime occurs when the mass
term $k^2$ in Eq.~(\ref{pert_kg}) becomes non-negligible. Therefore, we
assume that for all observable modes, the quintessence
perturbations reach the tracking solution before becoming
non-negligible; this simplifies the choice of initial conditions. In
fact it allows us to assume quintessence to be initially homogeneous.

%------------------------------------------------------------------------------------------------------------------------

\section{Supernova and CMB Likelihood Analysis}\label{like_res}
In this section, we present the results of the combined likelihood
analysis of the latest SN Ia and CMB data. Our goal is to identify
quintessence models characterized by a set of parameters ($\alpha,\Omega_m$), which are
statistically consistent with the observations (within $2\sigma$),
while departing from the standard $\Lambda \rm CDM$ values. 
These sets of parameters define what we call realistic quintessence models. 

SNe Ia standard candles provide a measurement of the
luminosity distance $d_L(z)$,
\begin{equation}
d_L(z)=(1+z)H_0^{-1}\int_0^z \frac{dz'}{E(z')},\label{dl}
\end{equation}
with $E(z)=H(z)/H_0$ is the dimensionless Hubble parameter and
$z=1/a-1$ is the cosmological redshift. Hence, the Hubble diagram
leaves $H_0$ unconstrained and degenerate in the plane
$\alpha-\Omega_m$ (Caresia, Matarrese \& Moscardini 2004; Schimd et al. 2007).
Therefore, we infer constraints from SN data after marginalizing
analytically the $\chi^2$ over $H_0$ (Lewis \& Bridle 2002; Di Pietro \& Claeskens 2004).
The CMB angular power spectrum is also degenerate in $\alpha-\Omega_m$
and $H_0-\alpha$ planes, however these degeneracies are
nearly orthogonal to those in the luminosity distance, thus providing
improved parameter bounds. In addition under the flat geometry
prior, CMB data allow for a precise measurement of $\Omega_m h ^2$ and
$\Omega_b h ^2$ (where $h=H_0/100$ km$/s$ $\rm Mpc^{-1}$), as well as the
primordial power spectrum properties (amplitude and slope).

\begin{figure}
\begin{center}
\begin{tabular}{c}
 \includegraphics[scale=0.4]{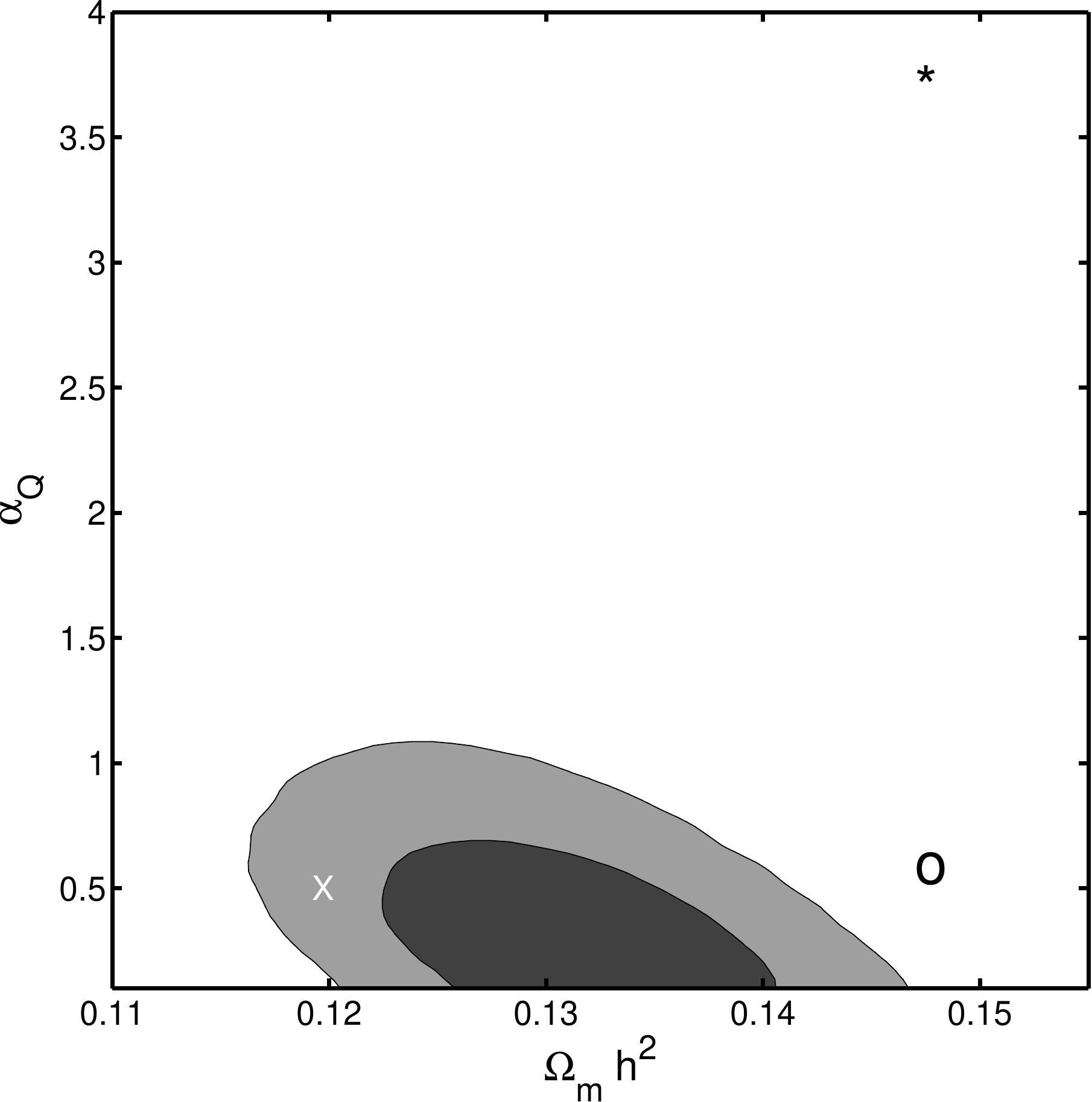} \\
 \includegraphics[scale=0.4]{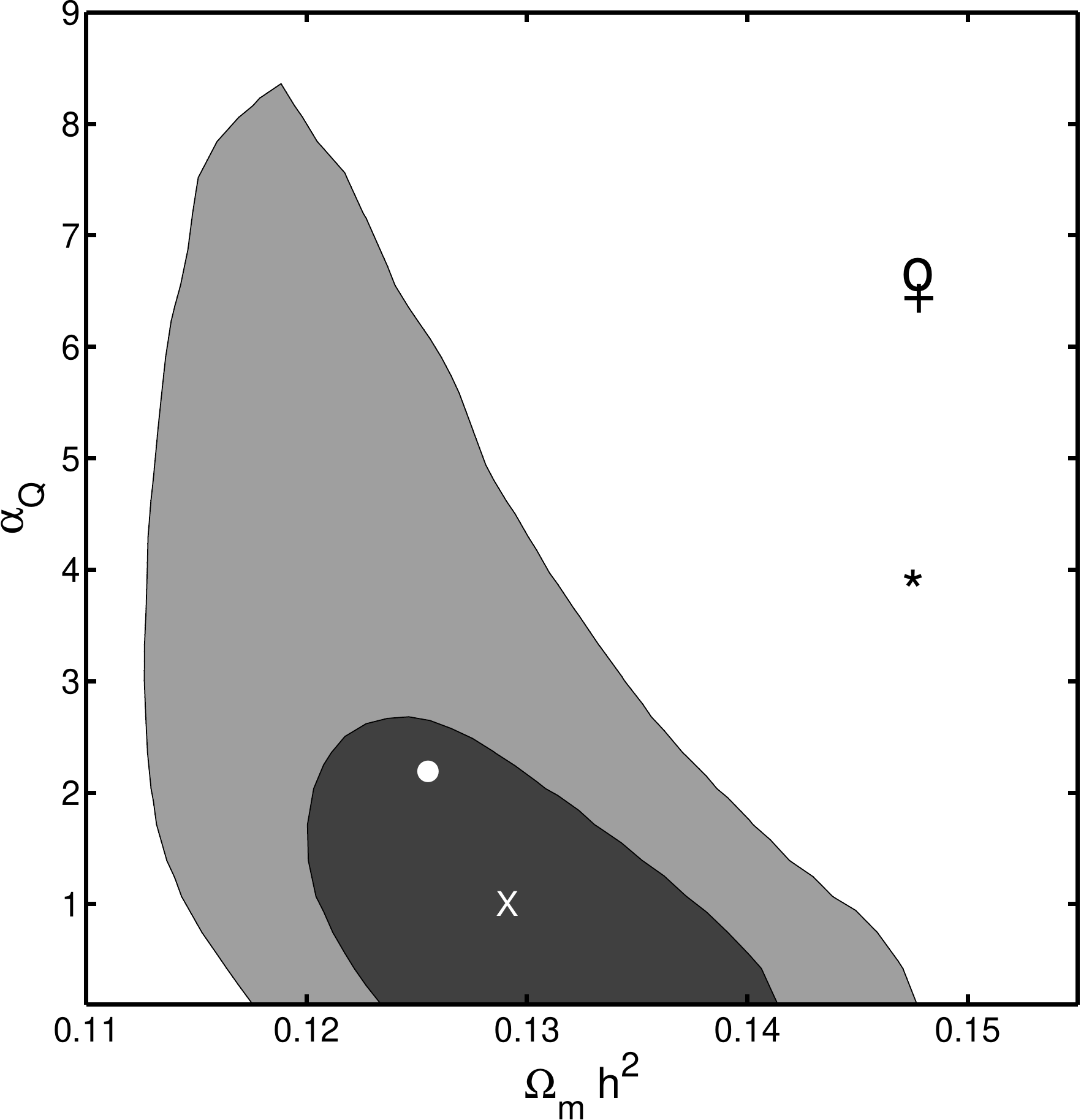}
\end{tabular} 
\end{center}
\caption{$68\%$ and $95\%$ confidence regions in the $\Omega_m h^2-\alpha$ plane from the combined analysis 
of the UNION SN Ia Hubble diagram and WMAP-5yrs CMB data for the Ratra-Peebles (top panel) and 
SUGRA (bottom panel). A cross mark (X) indicates our realistic quintessence model parameters 
choice, while model parameter values assumed in the literature are marked 
with * (Klypin et al. 2003 and Solevi et al. 2006), o (Dolag et al. 2004), + (Maio et al. 2006), $\bullet$ (Casarini et al. 2009).}
\label{conf}
\end{figure}

We perform a combined analysis of the WMAP5-yrs CMB power spectra
(Komatsu et al. 2008) and the recent compilation of SN Ia
measurements, UNION dataset (Kowalski et al. 2008). We run a series
of Markov Chains Monte Carlo using the publicly available codes \textsc{CAMB}
and \textsc{CosmoMC} (Lewis, Challinor \& Lasenby 1999; Lewis \& Bridle
2002), properly modified to correctly account for the physics
described in the previous section (influence of quintessence and
its perturbations). We assume uniform priors on the
following parameters: $\Omega_{b} h^2$, 
$\Omega_{\rm DM} h^2$, $h$, $\alpha$, $n_S$, $A_S$, $z_{rec}$, and
marginalize over $A_{SZ}$ (see Komatsu et al. 2008 for the exact definitions).
\\
\\
The results of the combined analysis are summarized in figure \ref{conf},
where we plot the $1$ and $2\sigma$ confidence regions in the plane
$\Omega_m h^2-\alpha$ for the RP (top panel) and SUGRA (bottom panel) models respectively.
As it can be seen the concordance $\Lambda \rm CDM$ model, which
corresponds to the limiting value $\alpha\rightarrow 0$ of the
quintessence parameter, is within the $1\sigma$ confidence region.
As noticed in previous quintessence/dark energy data analysis (see for
instance Corasaniti et al. 2004) quintessence models tend to fit the
data by requiring lower values of $\Omega_{m}h^2$, also confirmed by
the orientation of the degeneracy line in the $\Omega_m h^2-\alpha$
plane. As we will discuss more in detail in the next Section, 
this is because models with $\alpha>0$ are characterized by a
less accelerated cosmic expansion, therefore under the flatness requirement, they
require a larger amount of dark energy (lower matter density) to
account for the data. In Table \ref{tab1} we show the marginalized confidence
intervals on the model parameters for RP and SUGRA respectively. 
\footnote{It is worth noticing that supernova data do not provide
  strong constraints on slope and amplitude of the SUGRA
  potential. This is expected since the exponential term in
  Eq.~\ref{sugra} is dominant at late times, leaving these parameters
almost completely degenerate as shown by Caresia, Matarrese \&
  Moscardini (2004), and retrieved in Schimd et al. (2007).}

In figure~\ref{conf} we mark the model parameter values
assumed in previous works dedicated to the study of the effect of quintessence on
structure formation (Klypin et al. 2003 ; Dolag et al. 2004 ; Solevi
et al. 2006). These works have all assumed $\Omega_m=0.3$ and $h=0.7$,
corresponding to optimal fit parameter values of $\Lambda$CDM
cosmologies. Dolag et al. (2004) considered the case $\alpha=0.6$ for RP and $\alpha=6.7$ for
SUGRA (open circle), both leading today ($z=0$) to an equation of state
parameter $w_Q=-0.83$. Klypin et al. (2003) (see also Solevi et al. 2006) have assumed
the amplitude of the scalar potential to be $\lambda=1$ TeV, which for $\Omega_m=0.3$
corresponds to $\alpha\approx 4$ for both Ratra-Peebles and SUGRA
models (star) and gives a present eos $w_Q=-0.5$ and $w_Q=-0.85$, respectively.
Maio et al. (2006) have also considered a SUGRA potential and assumed
$\alpha=6.5$ (cross). It is evident that such models are ruled out at more
than $2\sigma$ by current CMB and SN Ia data. In addition these works
have assumed the same value of the total matter density as inferred in
a $\Lambda$CDM cosmology, hence it is not surprising that the authors
have concluded that in quintessence models the present structure formation process,
especially on the non-linear scales, slightly depart from the
$\Lambda$CDM scenario. An expection to the approach of these works is
the recent analysis by Casarini et al. (2009) which considered a Sugra model
with $\Omega_m=0.255$ and $\alpha_Q=2.2$ ($\bullet$ in Figure
\ref{conf}) corresponding to an eos value today $w_0=-0.908$
($h=0.7$), parameters inferred in another study by La Vacca et al. (2009).

In contrast we aim to study realistic models of quintessence which differ from the
$\Lambda$CDM, while being statistically consistent with current
cosmological data at level of significance that does not allow
them to be distinguished from the
$\Lambda$CDM. We choose a $\Lambda$CDM, RP and SUGRA
model specified by the parameter values given in Table \ref{tab2}. These 
are within the $2\sigma$ confidence limits, thus
differences of their $\chi^2$ values are not statistically significant.
The RP and SUGRA realistic model parameters are plotted in figure~\ref{conf}
as cross marks.
We want to stress that the inferred constraints on $n_s$, as well as
its best fit values are nearly similar for the three different
cosmologies (see Table \ref{tab1}), therefore without loss of generality we assume
$n_s=0.963$. For the Hubble parameter we assume
$h=0.72$, corresponding to the best fit value for a
$\Lambda$CDM cosmology (Komatsu et al. 2008). Although
the marginalized values of $h$ for RP and SUGRA models
are slightly lower (see Table \ref{tab1}), the value assumed is within the $95\%$
confidence interval. Such a choice allows us to directly compare
physical scales between different N-body simulations.
Hereafter $\Lambda \rm CDM$ will denote the concordance model, $\rm RPCDM$ quintessence
with Ratra-Peebles potential Eq. (\ref{rp}) and $\rm SUCDM$ quintessence
with SUGRA potential Eq. (\ref{sugra}). 

\begin{table}
\renewcommand{\arraystretch}{1.2}
\begin{center}
\begin{tabular}{ccc}
\toprule
Parameters &  $\rm RPCDM$ & $\rm SUCDM$\\
\midrule
$\Omega_{b}h^2$ & $0.0226\pm 0.0012$ & $0.0227\pm 0.0013$ \\
$\Omega_{\rm DM}h^2$ & $0.107^{+0.012}_{-0.011}$ & $0.105^{+0.012}_{-0.013}$ \\
$h$ & $0.674^{+0.047}_{-0.048}$ & $0.661^{+0.055}_{-0.058}$\\
$z_{reion}$ & $10.4\pm 2.7$ & $10.6\pm 2.8$\\
$ 10^{9} A_S$ & $2.12^{+0.18}_{-0.17}$ & $2.11^{+0.19}_{-0.17}$ \\
$n_s$ & $0.964^{+0.028}_{-0.029}$ & $0.968^{+0.034}_{-0.030}$ \\
$\alpha$ & $<0.89$ & $<6.2$ \\
\midrule
Derived parameters \\
$\Omega_{\rm m}h^2$ & $0.130\pm 0.012$ & $0.128\pm0.013$ \\
$\Omega_m$ & $0.287^{+0.056}_{-0.050}$ & $0.293^{+0.060}_{-0.053}$ \\
$\sigma^{\mathrm{lin}}_8$ & $0.737^{+0.083}_{-0.097}$ & $0.70^{+0.11}_{-0.15}$ \\
$\log_{10}\lambda (GeV)$ & $-8.9^{+3.3}_{-2.0}$ & $-3.7^{+12.5}_{-6.8} $ \\
$w_0$ & $<-0.78$ & $<-0.83$\\
$w_1$ & $<0.11$ & $<0.39$\\
$q_0$ & $0.44^{+0.13}_{-0.17} $ & $0.47^{+0.13}_{-0.16}$ \\
$z_{\rm acc}$ & $0.67^{+0.15}_{-0.16}$ & $0.58^{+0.18}_{-0.19}$ \\
$z_{\rm d}$ & $0.42^{+0.12}_{-0.11}$ & $0.40^{+0.12}_{-0.11}$ \\
$t_0\ (Gyr)$ &$13.84^{+0.28}_{-0.27}$ & $13.95^{+0.38}_{-0.32}$ \\
\bottomrule
\end{tabular}
\caption{Cosmological parameters and their 95\% confidence level
  intervals obtained from the combined analysis of WMAP5 and UNION SN
  Ia data. $w_0$ is the present value of the DE eos and $w_1$ is the
  first order parameter of the linear eos parameterization
  $w(a)=w_0+w_1(a-1)$ (Chevallier \& Polarski, 2001 ; Linder, 2003, quintessence only allows $w_0>-1$ and $w_1>0$), $q_0$ is the
  present value of the acceleration factor, $z_{\rm acc}$ is the
  redshift marking the beginning of cosmic acceleration, $z_{\rm
  d}$ is the redshift of DE domination), $A_S$ is the primordial amplitude of curvature perturbations at $k=0.05 \text{Mpc}^{-1}$.} 

\label{tab1}
\end{center}
\end{table}

\begin{table}
\begin{center}
\begin{tabular}{cccc}
\toprule
Parameters  & $\Lambda \rm CDM$ & $\rm RPCDM$ & $\rm SUCDM$\\
\midrule
$\Omega_{m}$ & $0.26$ & $0.23$ & $0.25$ \\
$\alpha$ & $0$ & $0.5$ & $1$\\
$A_S$ & $2.1\times 10^{-9}$&$2.0\times 10^{-9}$ &$2.1\times
10^{-9}$\\
\midrule
Derived parameters \\
$\sigma_8^{lin}$ & $0.80$ & $0.66$ & $0.73$ \\
$\lambda$(eV) & $2.4\times 10^{-3}$ & $4.9$ & $2.1\times 10^{3}$\\
$w_0$ & $-1$ & $-0.87$ & $-0.94$\\
$w_1$ & $0$&$0.08$ &$0.19$\\
\bottomrule
\end{tabular}
\caption{Cosmological parameters selected for the realistic
  models. These are flat models ($\Omega_{Q(\Lambda)}=1-\Omega_{m}$),
  with a spectral index $n_s=0.963$, $h=0.72$, $\Omega_b h^2=0.02273$, and
  $\tau=0.087$ corresponding to $z_{reion}=10.1$ for RPCDM and
  $z_{reion}=10.4$ for SUCDM respectively. } 
\label{tab2}
\end{center}
\end{table}

%%%%%%%%%%%%%%%%%%%%%%%%%%%%%%%%%%%%%%%%%%%%%%%%%%%%%%%%%%%%%%%%%%%%%%%%%%%%%%%%%%%%%%%%%%%%%%%%%%%%%%%%%%

\section{Phenomenology of realistic quintessence models}\label{quinte_impr}
Here we describe the main features of the realistic quintessence cosmologies
which we inferred from the likelihood data analysis described in the previous section
and for which we aim to study the non-linear structure formation through N-body simulations. 
Although the characteristics of the background evolution and the linear density perturbations 
for the RP and SUGRA models, and more in general time varying dark energy equation of state models 
have been studies in a vast literature, we think that the reader may benefit from a 
brief review of how the scalar field dynamics impact the cosmic expansion and 
the evolution of the matter perturbations. In fact this will allow for a necessary 
understanding of the effects that dark energy leaves on the
non-linear structure formation which will be discussed in the next section.

Quintessence affects the linear clustering of matter 
through the combined effect on the background evolution and
the presence of dark energy perturbations. 

\begin{figure}
\begin{center}
\begin{tabular}{c}
\includegraphics[scale=0.4]{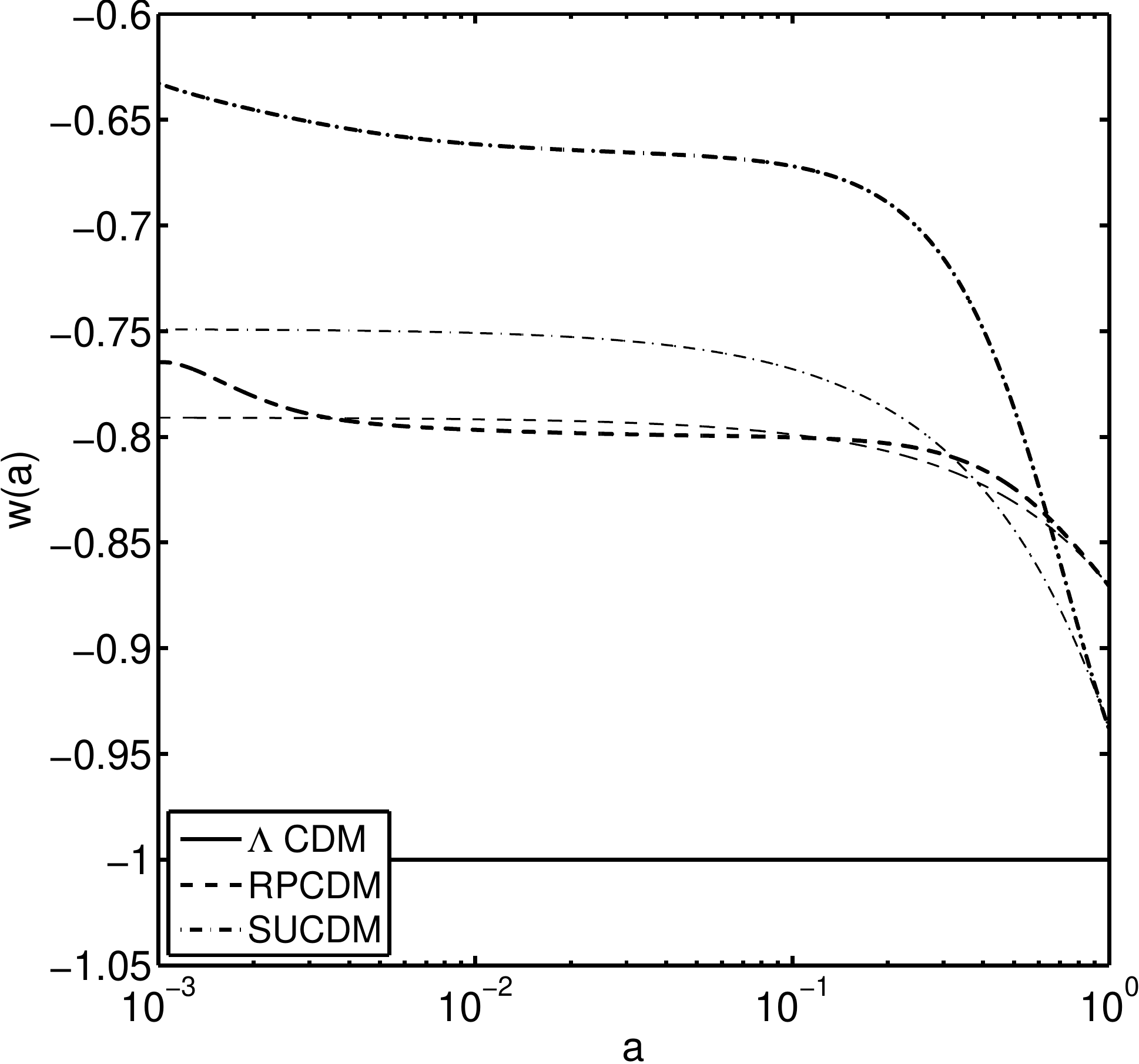} \\
\includegraphics[scale=0.4,angle=0]{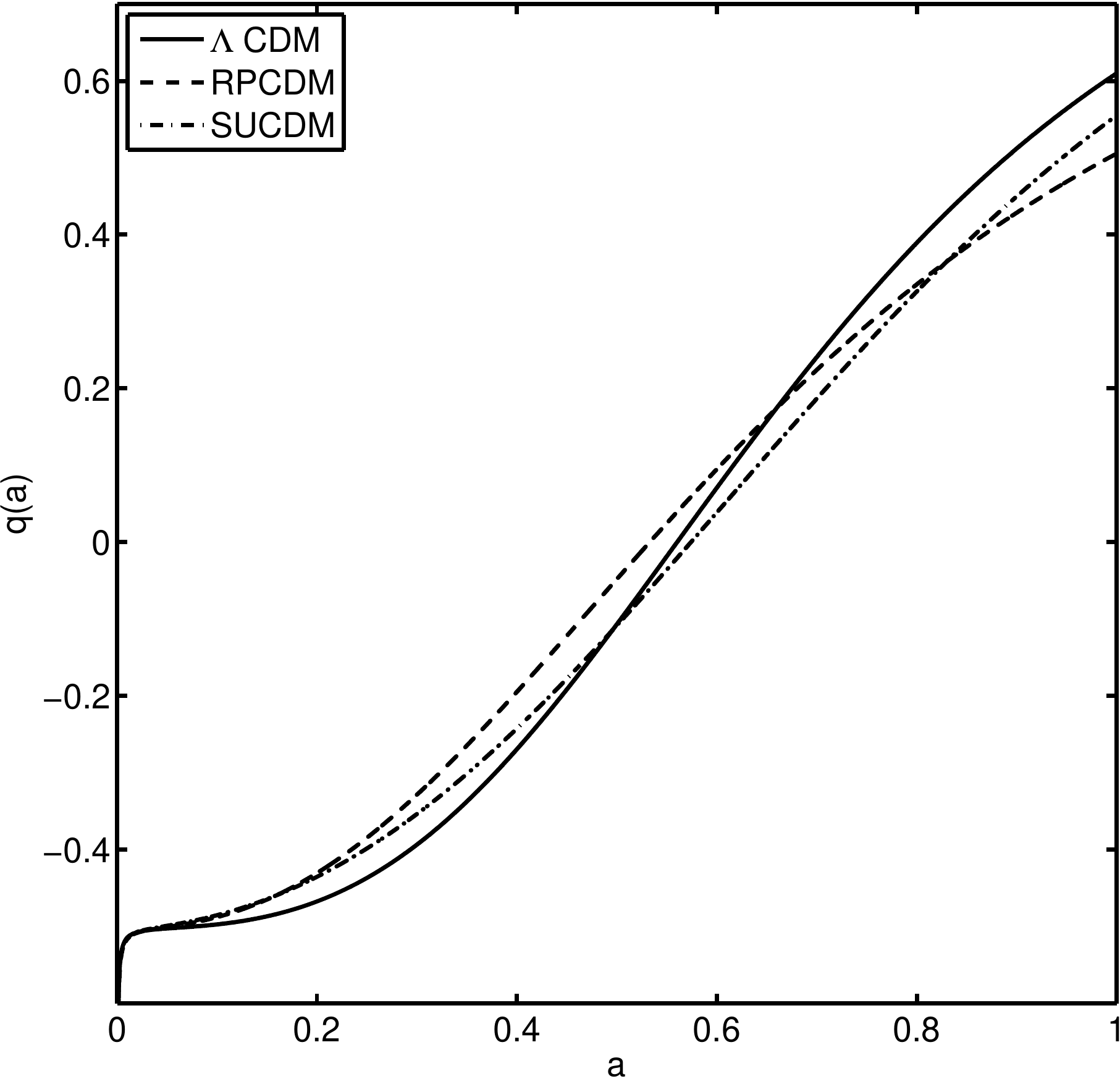} 
\end{tabular}
\end{center}
\caption{Evolution of the equation of state
  $w=p/\rho$ (top panel) and acceleration factor $q=\mathcal{H}'/\mathcal{H}^2$ (bottom panel)
as function of the scale factor $a$ for the models in Table
  \ref{tab2}. For comparison we also plot the linear eos parametrization for each
  model. 
}
\label{wq}
\end{figure}

\begin{figure}
\begin{center}
\includegraphics[scale=0.4]{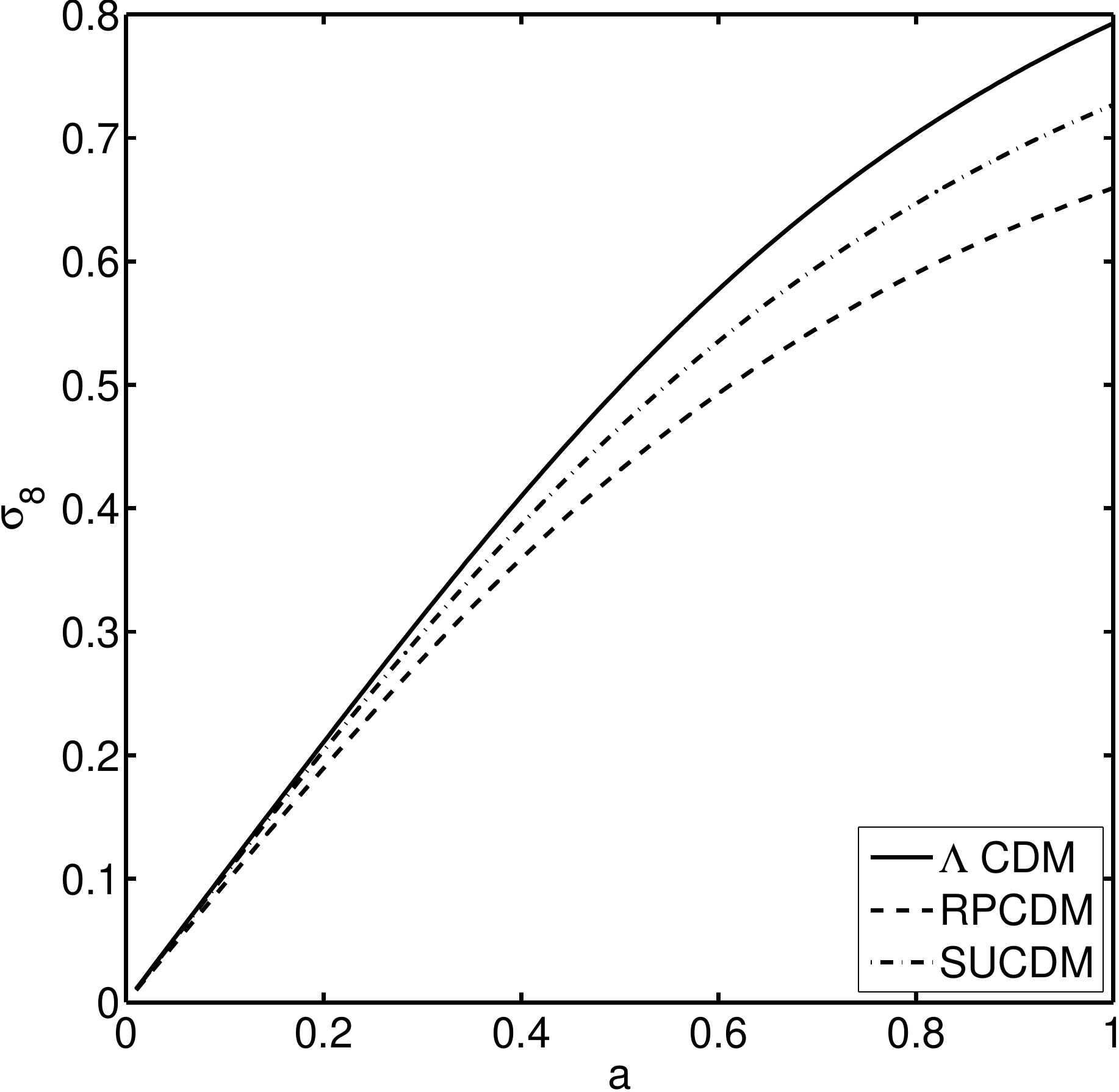} \\
\includegraphics[scale=0.4]{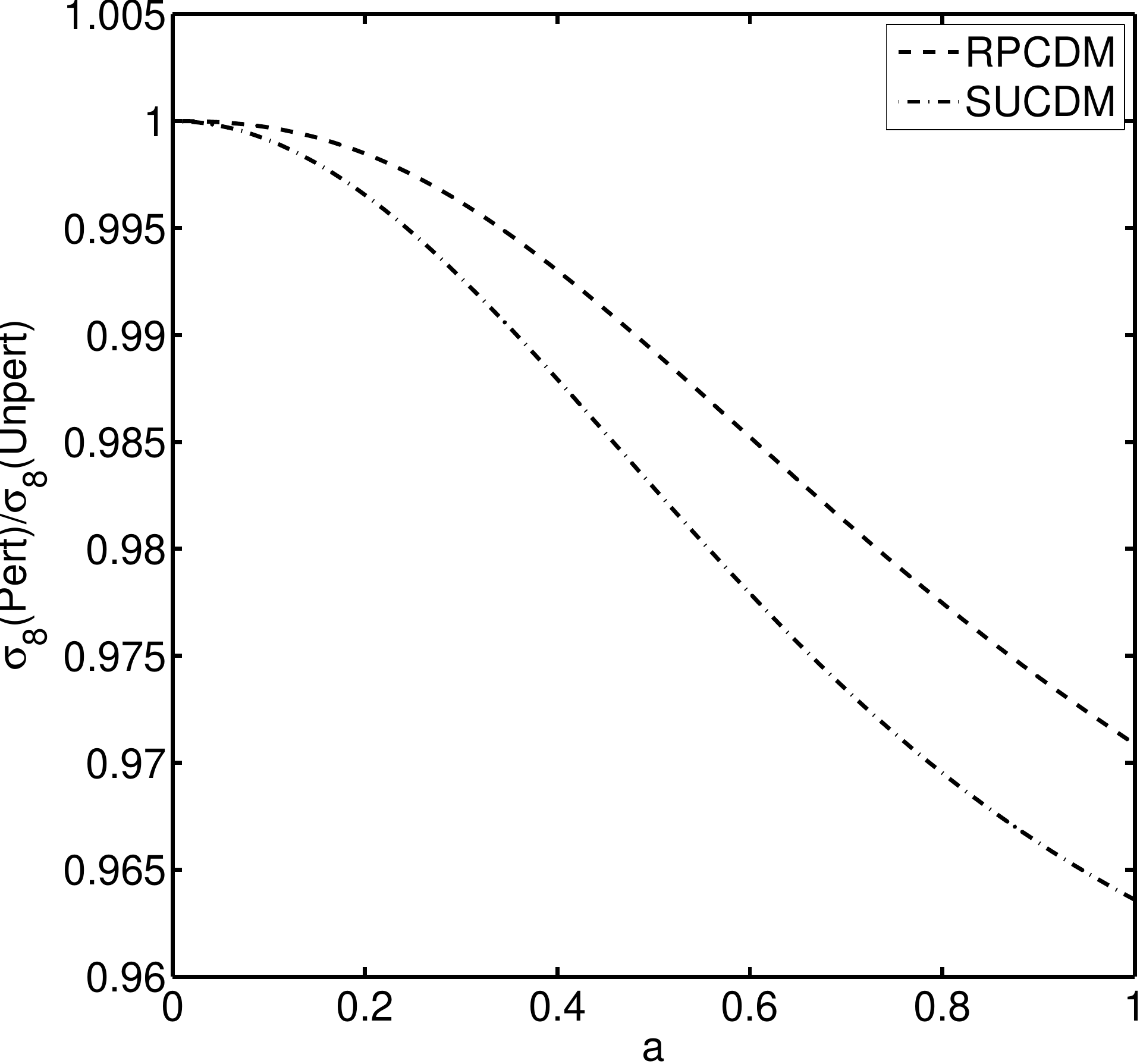}
\end{center}
\caption{Evolution of the $\sigma_8$ parameter for the models in Table \ref{tab2} (top). 
Reshift evolution of the ratio of the $\sigma_8$ value with and without quintessence perturbations (bottom).
}
\label{sigma}
\end{figure}

\begin{figure}
\begin{center}
\includegraphics[scale=0.4]{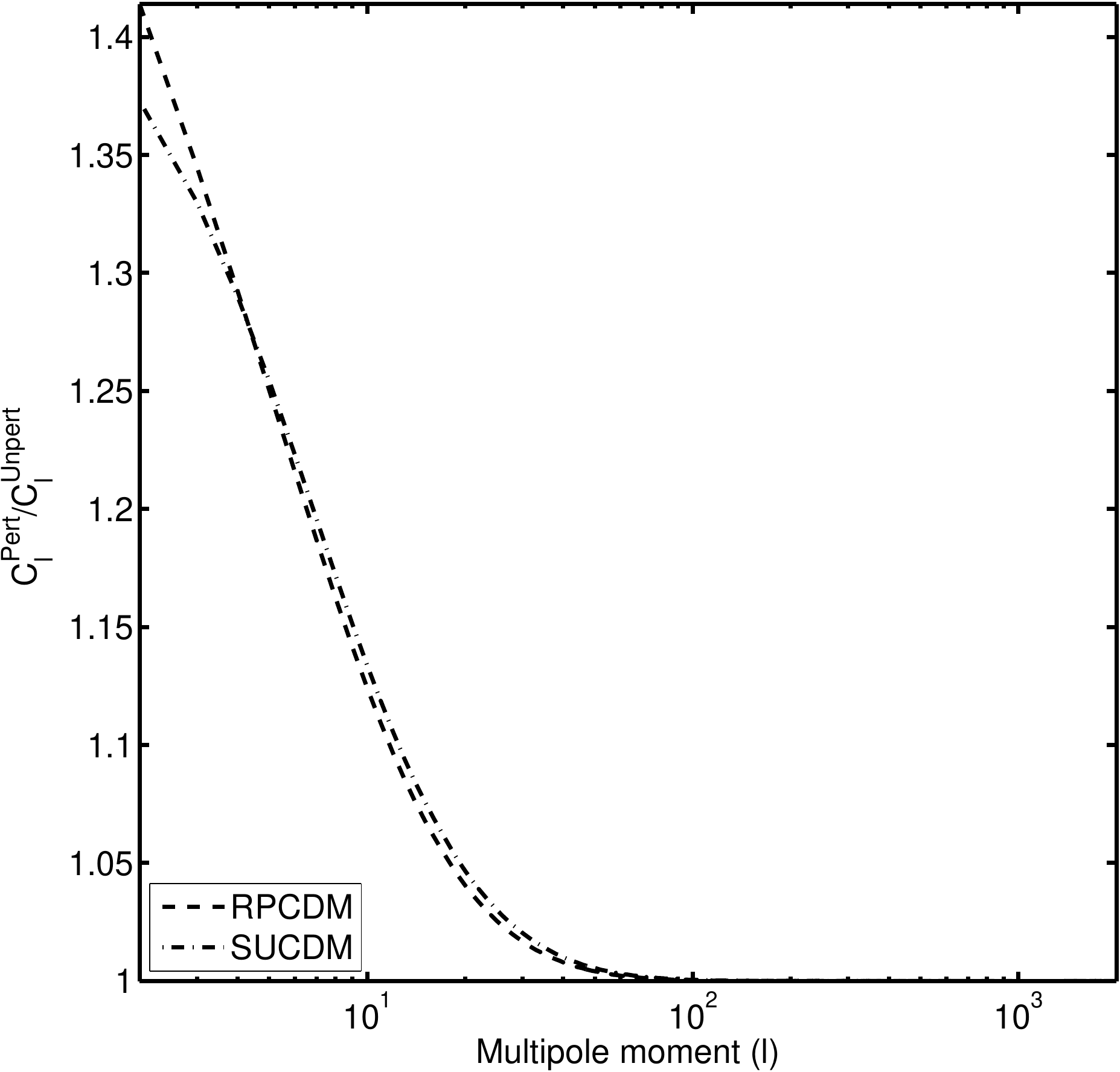}
\caption{Ratio between the angular power spectrum of the CMB anisotropies with
  and without quintessence perturbations for the models in Table \ref{tab2}.}\label{c2l}
\end{center}
\end{figure}

\begin{figure}
\begin{center}
\begin{tabular}{c}
\includegraphics[scale=0.4]{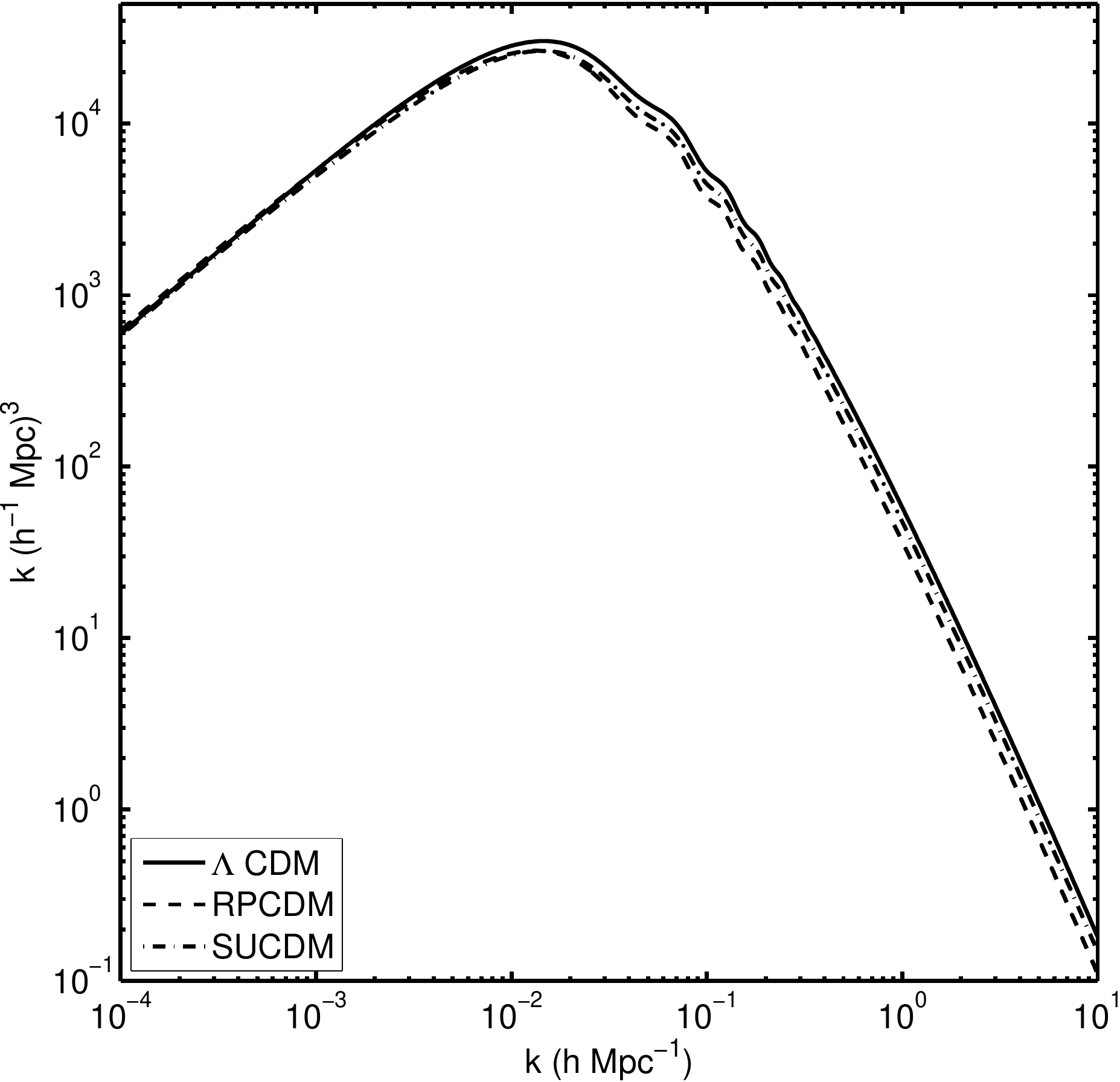} \\
\includegraphics[scale=0.4]{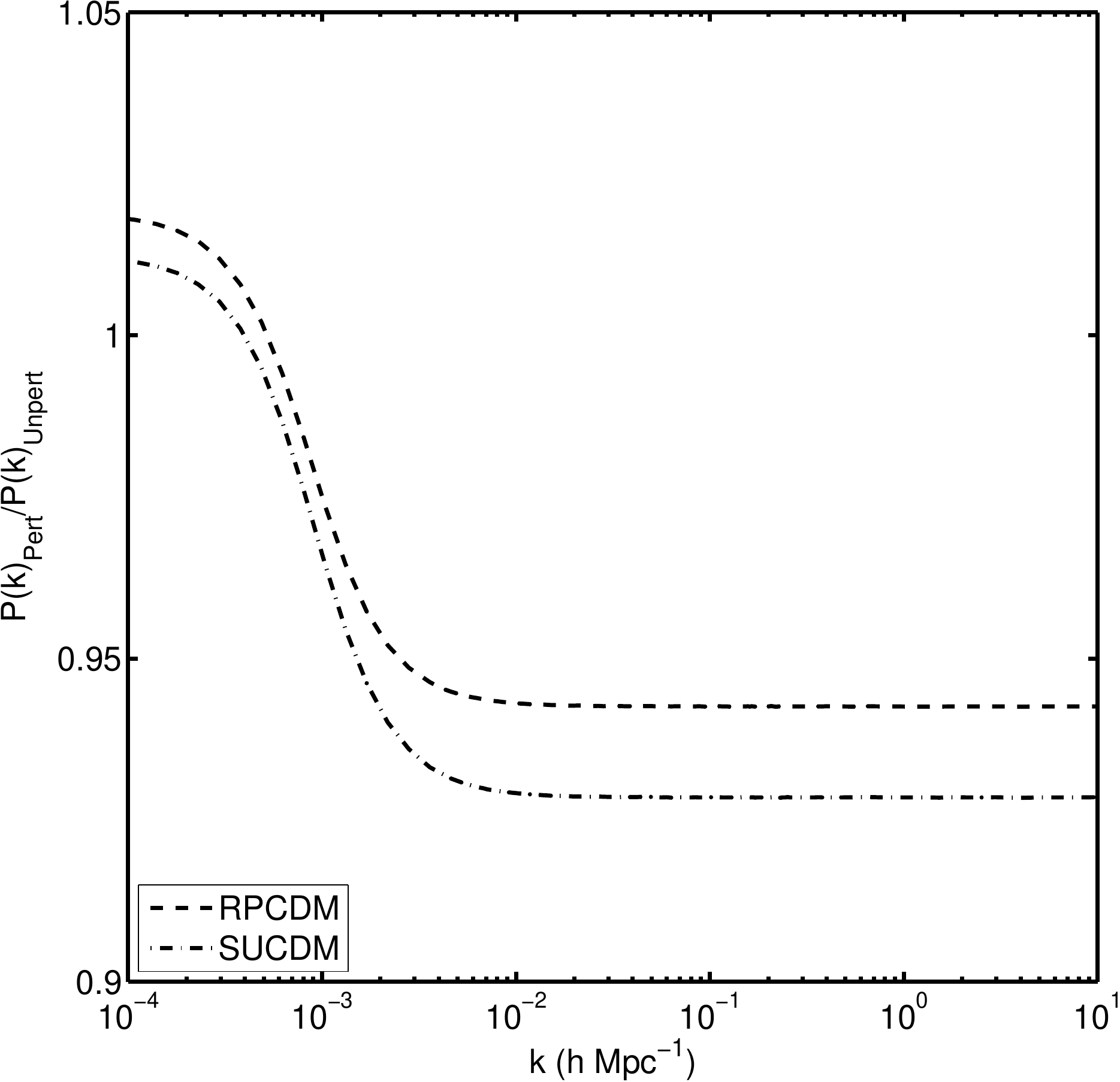}
\end{tabular}
\caption{CMB-normalized linear matter power spectra for the
  models in Table \ref{tab2} (top panel). Ratio between the power spectrum with
  and without quintessence perturbations (bottom panel).}\label{pks_lin}
\end{center}
\end{figure}

Let us first focus on the background dynamics of quintessence models. 
In the top panel of figure~\ref{wq} we plot the
evolution of the quintessence eos as a function of the scale factor for
the RPCDM (dash line) and SUCDM (dash-dot line) models, 
and the $\Lambda$CDM (solid line) respectively. We can see that these curves are
characterized by different behaviors which are indicative of the
different cosmic expansion histories. In particular we may notice that
during the matter dominated era the quintessence eos is nearly constant
(consistently with the fact that the field is in the tracker solution),
with the SUCDM eos larger than the RPCDM case, while at later times it rapidly
evolves with the SUGRA eos becoming more negative than in the RP case.
This is because at early times the field is evolving over an inverse power law potential
and since the RP model considered is flatter than SUGRA (i.e. the slope $\alpha_{RP}<\alpha_{SUGRA}$),
hence $w_{RP}<w_{SUGRA}$ during the matter dominated era, and the flatter the potential  
the smaller is the variation of the equation of state. 
In contrast at late times (when $\varphi\approx 1$) in the SUGRA model the field rolls
over a region of the where the contribution of the
exponential term in Eq.~(\ref{sugra}) is important, thus making the curvature smaller
than that of RP, thus the kinetic energy rapidly
decreases causing the present SUGRA eos value to be more negative than 
the RP case. In figure~\ref{wq} (top panel) we also plot the linear
eos parameterization, $w(a)=w_0+w_1(1-a)$ (Chevallier \& Polarski
2001; Linder 2003), which are commonly used in the literature to 
mimic the quintessence model dynamics. We may notice that such
parameterization provide a valid approximation at low redshift, on the
other hand during the matter dominated era it can account for the quintessence evolution 
only reasonably well for very flat potentials. More
specifically, the accuracy of this parameterization at $z\approx 1000$ varies from about
$4\%$ for the RPCDM to $15\%$ for the SUCDM.
This specific behaviours of the quintessence equation of state 
affects the cosmic expansion history, as it can be seen in the bottom
panel of figure~\ref{wq} where we plot the acceleration
factor $q=\mathcal{H}'/\mathcal{H}^2$ for the three different realistic models. 
Deep in the radiation dominated $q=-1$ for all models, while the
subsequent evolution during the matter-dominated era differs. We can see that
before the deceleration-to-acceleration transition takes place (corresponding to $q$ changing sign), 
$|q_{\Lambda\rm CDM}|>|q_{QCDM}|$, thus indicating that the $\Lambda$CDM model
decelerates more than in realistic quintessence cosmologies. The acceleration
occurs nearly at the same redshift for all models, although for the specific
model parameter choice we have that RPCDM accelerates slightly earlier
than the $\Lambda$CDM and SUCDM. On the other hand for $a>0.6$ we have
$q_{\Lambda\rm CDM}>q_{QCDM}$, hence the acceleration in $\Lambda$CDM   
is stronger than in quintessence models, which implies that 
in order to fit the data as well as the $\Lambda$CDM, the smaller 
acceleration of the quintessence cosmologies is compensated by
a larger amount of dark energy density, i.e. (in a flat universe) 
a smaller amount of matter. This is indeed the case, since 
for our realistic models for which we have $\Omega_m=0.23$ 
for the RPCDM and $\Omega_m=0.25$ for the SUCDM while $\Omega_m=0.26$ in $\Lambda\rm CDM$ (see Table \ref{tab2}).
The different cosmic expansion histories affect 
the linear evolution of the matter density perturbations in a rather peculiar
way resulting in a suppression of their growth rate at late times.
This can be understood by consider the equation for the
linear growth factor of the matter density perturbations, $D_+(a)$:
\begin{equation}
\label{dlin}
\dfrac{d^2 D_+}{da^2}+\left(\dfrac{d\ln \mathcal H}{da}+\dfrac 2 a\right)\dfrac{d D_+}{da}-\dfrac{3 \Omega_m H_0^2}{2 a^3 \mathcal H^2} D_+ = 
\dfrac{1}{a^2\mathcal H^2 T_m(k)} \left( \dfrac{a^2 \delta\varphi}{m_{Pl}^2} \dfrac{dV}{d\varphi} -2 a^2 \mathcal H^2 \dfrac{d\varphi}{da}\dfrac{d\delta\varphi}{da}\right)
\end{equation}
where $T_m(k)$ is the matter transfer function. Since we aim to isolate 
the effect of the background evolution from that of the dark energy perturbations
let us neglect for the moment the feedback of the quintessence field fluctuations, and set to zero
the right-hand-side of Eq.~(\ref{dlin}). We may notice that in such a case the evolution
of the growth factor becomes scale-independent. In figure~\ref{sigma} (top) we
plot the normalisation of the linear matter power spectrum (through the $rms$ mass fluctuation amplitude  $\sigma_8(a)$ in spheres of size $8h^{-1}$ Mpc) 
for the three realistic models. As already mentioned the growth
factor in quintessence cosmologies is suppressed at late times
compared to the $\Lambda$CDM case. As can be seen in
figure~\ref{sigma} (bottom), on such a scale the presence of quintessence
perturbations increases the power suppression by a few percent at $z=0$
(see also figures \ref{c2l} and \ref{pks_lin} for the spatial dependence of this effect).

\begin{figure}
\begin{center}
\includegraphics[scale=0.4]{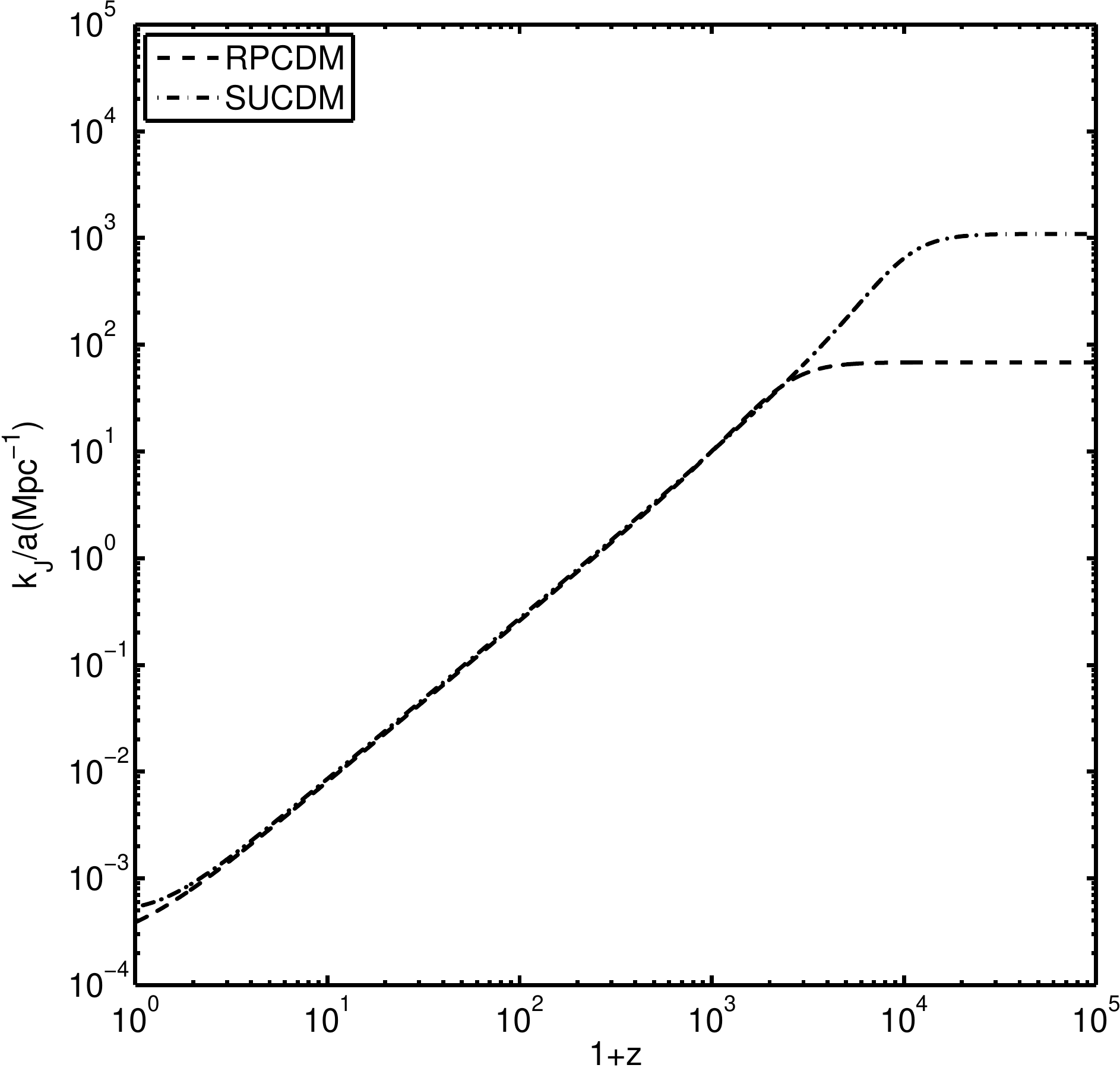} 
\caption{Reshift evolution of the physical Jeans wavenumber $k_J/a=\frac{1}{m_{Pl}}\left(\frac{d^2V}{d\varphi^2}\right)^{1/2}$ associated 
to the quintessence fluid.}\label{jeans}
\end{center}
\end{figure}

In fact quintessence perturbations cause an additional scale dependent effect on the
linear clustering of matter. A simple way to evaluate such a scale dependence is
to consider quintessence as a fluid with comoving Jeans-mode 
given by the curvature of the potential, i.e. the mass of the field, 
$k_J=a/m_{Pl}\sqrt{d^2V/d\varphi^2}$. Therefore, scales
which correspond to modes $k<k_J$ will collapse under gravitational instability,
while modes $k>k_J$ will undergo a series of damped
oscillations due to pressure waves in the quintessence fluid. Since
quintessence is a light field, by the time it becomes the dominant energy
component its clustering has halted on the small scales, while allowing for some 
level of clustering on the larger ones. In figure~\ref{jeans} we plot the redshift
evolution of $k_J$ for the RPCDM and SUCDM models respectively. We can 
see that at $z=0$ the Jeans-scale correspond to $k_J\approx 3\times 10^{-4}$Mpc$^{-1}$,
hence today only quintessence perturbations which are on the very large
scales can cluster.

This has two major consequences. Firstly, the large scale clustering of dark energy
enhances the amplitude of the ISW effect
on the CMB power spectrum (Corasaniti et al. 2003; Lewis and Weller 2003), 
as it can be seen in figure~\ref{c2l} where we plot the ratio of the
CMB temperature power spectrum of the realistic models with and without quintessence perturbations. 
Secondly, on the small scales where
dark energy is nearly homogeneous, the growth factor is suppressed due to the effect
of the background dynamics. Therefore in combination with the power
enhancement on the large scales caused by the dark energy
perturbations, the overall effect of quintessence on the linear matter power 
spectrum is to produce a different distribution of power between the
small and large scales with respect to the $\Lambda$CDM scenario.

This can be seen in figure~\ref{pks_lin} where we plot the
CMB normalized linear matter power spectra
(top panel), and the ratio of the linear matter power spectra with and
without quintessence perturbations (bottom panel) for the realistic
models. From the latter we can clearly see that on the very large
scales ($k<k_J$) the dark energy clustering
enhances the matter power spectrum compared to the unclustered case, while on
small scales ($k>k_J$) the opposite occurs. This is consistent with
the results of Ma et al. (1999), where the authors showed that
for increasing values of a constant eos the amount of dark matter 
clustering on the large scales is enhanced, while on small scales it
remains unchanged when compared to the $\Lambda$CDM case. 
Consequently when considering CMB normalized 
spectra, quintessence models generally predict smaller values
of the $\sigma_8$ than the $\Lambda$CDM as shown by Kunz et al. (2004)

Indeed not including the effect of the dark energy perturbations would
lead to overestimated values of $\sigma_8$; for instance
assuming a homogeneous quintessence we find $\sigma_8=0.68$ for
the RPCDM model (instead of $0.66$) and $\sigma_8=0.76$ (instead of
$0.73$), corresponding to a loss of about $3\%$ and $4\%$ for RP and
SUGRA respectively (see also bottom panel of figure \ref{sigma}). 
Accounting for dark energy perturbations requires solving Eq.(\ref{dlin}) with a non-vanishing rhs,
and results in a lowering of the homogeneous solution ($\delta\varphi=0$) for the linear growing mode $D_+(a)$.

We find the WMAP normalized linear matter power spectra, when evaluated including
the effect of dark energy clustering, are statistically consistent 
with the power spectrum estimates from the Sloan Digital Sky Survey 
data (SDSS, Tegmark et al. 2006). For consistency we limit to the reduced 
dataset including point up to $k=0.2 \rm h^{-1} Mpc$ which are not 
affected by non-linear corrections. The three realistic models best
fit the SDSS data with $\chi^2$ differences smaller than unity, 
provided the galaxy bias parameter, $b_g$, assumes the following
values: $b_g=2.02$ for RPCDM, $b_g=2.01$ for SUCDM and $b_g=1.98$ for $\Lambda$CDM.
 
It is worth remarking that the effects due to the background evolution
and clustering of quintessence are strictly related to the features of
the quintessence potential. Although the scalar potential can be expressed in terms
of time derivatives of the equation of state parameter (see e.g. Dave
et al. 2002), the simplest 
eos parametrizations used in previous works can at most 
reproduce the quintessence dynamics only at very small redshift, 
failing to describe features related to the scalar potential 
and its higher derivatives at high redshift. Such features can in principle
be accounted for only when using more complicated multi-dimensional 
parametrizations. Therefore, the results of N-body simulations of dark energy
models specified by a constant or a linearly scale factor evolving equation
of state are intrinsicaly incomplete with respect to the fundamental 
scalar field evolution approach that we have developed in this study.

\section{The Non-linear Matter Power Spectrum}\label{quinte_sim}

In this section, we present the N-body simulations that we have
performed to study the imprints of realistic quintessence cosmologies
on the non-linear structure formation. We will first introduce the
numerical methods developped, then we will describe the characteristics of the 
cosmological simulations and finally discuss the results.

\subsection{Initial conditions}

The initial conditions for the numerical simulations have been calculated using the MPGRAFIC
package described in great details in Prunet et al. (2008). This code is a parallel version of GRAFIC
(Bertschinger 1995; Bertschinger 2001) and generates gaussian random
fields\footnote{We however note that MPGRAFIC does not use any Hanning filter unlike the default version of GRAFIC. This is important, especially for large scale simulations,
because the Hanning filter damps out small scale modes.}. Initial displacements and
velocities of the dark matter particles are computed using the Zel'dovich approximation at early cosmic time $\tau$,
\begin{align}
\textbf{x}(\textbf{q},\tau)&=\textbf{q}+\frac{D_+(\tau)}{D_+^0}      \textbf{d}(\textbf{q}),\\
\textbf{v}(\textbf{q},\tau)&=           \frac{\dot{D}_+(\tau)}{D_+^0} \textbf{d}(\textbf{q}),
\end{align}
where $\textbf{x}$ is the perturbed comoving position, $\textbf{v}$ is the
proper peculiar velocity, $\textbf{q}$ is the Lagrangian coordinate corresponding to the
unperturbed comoving position. The dot denotes a derivative with respect to $\tau$ and $ \textbf{d}(\textbf{q})$ is the
displacement field computed from the density fluctuation field.  The
latter is obtained as convolution of a random white noise (which defines the phase of the
realization of the Universe) with the square root of the linear power spectrum (which defines
the amplitude of the fluctuations at different scales).

The public version of MPGRAFIC includes dark energy only as a
cosmological constant, in order to adapt the code to quintessence
cosmologies, we have first input MPGRAFIC with the power spectrum 
at z=0 computed with the modified version of \textsc{CAMB} (described in the previous
section). Then, we have modified all the cosmological routines (such as $a(\tau)$,
$D_+(\tau)$ and $\dot{D}_+(\tau)$) to account for the quintessence effects. 
In particular, the growth
factor routine was modified  to account for Eq. (\ref{dlin}) (with homogeneous quintessence and a vanishing right-hand-side
as explained above and in section 5.2). In those quintessential universes, DE dominates lately
in expansion history (at $z_d<0.45$ see Table 1) so that the initial conditions for $D_+(a)$ for very small $a$ correspond
to matter dominated Einstein-deSitter cosmology\footnote{In practice, we chose $D_+(a_i=0.002)=0.6 a_i$ and 
$(d D_+/d a)|_{a_i}=0.6$ (see also Peebles, 1993). Note that only the quantity $D_+(a)/D_+(a_0=1)$ matters in the normalisation
of cosmological initial conditions.}. We have cross-checked the obtained growth factor with the solution computed by the \textsc{CAMB} code
in the case of vanishing DE perturbations.
At
this point, some important remarks are necessary concerning the
generation of the initial conditions. For each cosmological simulation
we have considered the same realization of the Universe, that is to
say we have used the same white noise (more specifically the Horizon white
noise\footnote{http://www.projet-horizon.fr}). This choice 
allows us to properly compare the results between different
models. As far as the choice of the initial
redshift is concerned, we started all simulations with the same amount of
fluctuations at the scale of the grid resolution which is given by $\sigma(L/n, z_{i})$, where
$L$ is the box length and $n$ is the number of coarse grids along one spatial
direction. Again this requirement allows us to consistently compare
different models. The initial redshift $z_i$, then depends on the cosmology and
box length considered, and is determined by solving the equation
$$
\frac{\sigma(L/n, z_{i})}{\sigma(L/n, z=0)}=\frac{D_+(z_i)}{D_+(z=0)},
$$
where $D_+(z)$ is the linear growing mode obtained solving Eq. (\ref{dlin}) without quintessence perturbations ($\delta\varphi=0$).
As we will discuss more clearly in the next section this is because the N-body simulation code does not account for the
gravitational collapse of quintessence (see section 5.2 below). On the
other hand $\sigma(L/n, 0)$ is obtained using the linear matter power spectrum at
$z=0$ calculated including the effect of quintessence perturbations (see sections 4 and 5.2),
and $\sigma(L/n, z_{i})$ is fixed to a small arbitrary value.
We choose $\sigma(L/n, z_{i})=0.05$, for which we expect small
artificial effects due to the Zel'dovich approximation to be small.
In fact, Crocce et al. (2006) by using an accurate second-order
Lagrangian Perturbation Theory (2LPT) to set-up the initial conditions,
have shown that a late time start of the N-body simulations using the Zel'dovich
approximation tends to underestimate the non-linear power spectrum at
the level of $2\%-8\%$ (depending on the redshift). In their analysis the initial redshift
of the simulations is set to $z_i=49$. In our case, by using the above
method and considering the same cosmology and simulation
box length we have $z_i=72$, which is high (and even much higher than $z_i=30$
used in Smith et al. 2003). Finally it is worth stressing that our
analysis mainly focus on the difference between models. Therefore, spurious
effects caused by the Zel'dovic approximation are expected to be
negligible, and unlikely to affect our main conclusions.

\subsection{Cosmological code}

The N-body simulations have been performed using the \textsc{RAMSES} code (Teyssier 2002; Rasera \& Teyssier 2006), 
which is based on an Adaptive Mesh Refinement (AMR) technique, with a tree-based
data structure that allows for recursive grid refinements on a cell-by-cell
basis. Particles are evolved using a particle-mesh (PM) solver, while the Poisson equation is 
solved by a multigrid method. The refinement strategy follows a ``quasi-lagrangian'' approach
where cells are divided by 8 if their enclosed masses are multiplied
by 8. We did not impose any maximum level of refinement for the
simulations, and let the code trigger as much refinement levels as needed. This allows us to have very high
resolution down to the core of dark matter halos (up to $2.5$ h$^{-1}$kpc).
\textsc{RAMSES} has been parallelized using a dynamical domain decomposition based on the
Peano-Hilbert space-filling curve. This is an important
improvement of the code for high-resolution runs with a high level of
structuration. In order to implement the cosmological evolution of the different
models we have modified all cosmological routines such as to render
them model-independent. These are now input with pre-computed numerical tables of cosmological 
quantities needed to fully specify the cosmological model evolution. In order to check that the overall
implementation is correct, we compare the evolution of the power
spectrum on the large scales ($>500$~Mpc) to the prediction of the linear calculation
using the \textsc{CAMB} code and find an agreement at the
percent level. 

Ideally, the gravitational collapse of quintessence should be
incorporated in the N-body code. However, the collapse of a
negative pressured quintessence fluid is a very complicated task that will
require, in addition, a considerable amount of numerical resources
to the detriment of precision on the matter fluctuations.
A naive approach would consist of linearly propagating the
matter power spectrum starting from some initial conditions set deep in the
radiation era till the initial redshift of the simulation. Nonetheless
using such an approach would completely miss the late time influence of quintessence
perturbations, thus failing to reproduce the correct linear matter power
spectrum at $z=0$ at least on those large scales that evolve linearly
in the simulation box. A better approximation is realized 
by using the linear matter power spectrum at $z=0$ computed from the
linear perturbation theory accounting for the presence of quintessence perturbations
(for instance using the \textsc{CAMB} code), then evolve backwards such a spectrum
to the starting redshift of the simulation by using
the linear growing mode $D_+(a)$ evaluated without including the quintessence perturbations.
In such a case, we are guaranteed by construction that 
the evolution of dark matter particles on the scales that 
remains linear in the N-body simulation recovers 
the matter power spectrum at $z=0$ when
quintessence perturbations are included. An ideal quintessence
simulation would account for more subtle effects, such as a coupling
between the non-linear evolution of DM on the small scales and the
linear clustering of DE on the large ones, as well as a coupling
between the non-linear clustering of DM and that of DE on the small
scales. Developing such a simulation requires a study that goes beyond
the scope of our analysis and we leave it for future work.
Nevertheless, we think that our approach by including the effect of the
large scale linear evolution of DE perturbations is an important
improvement compared to previous N-body simulation studies which have
implemented dark energy only through its effects on the background evolution.

\begin{figure}
\begin{center}
\begin{tabular}{ccc}
\includegraphics[scale=0.59]{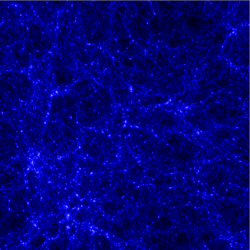} &
\includegraphics[scale=0.59]{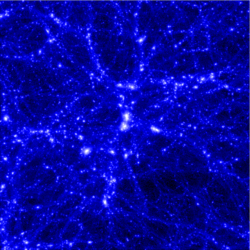}&
\includegraphics[scale=0.59]{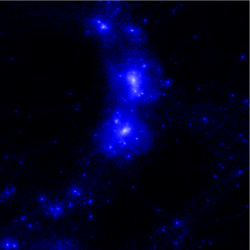} \\
\includegraphics[scale=0.59]{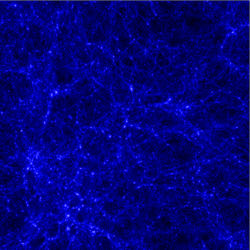} &
\includegraphics[scale=0.59]{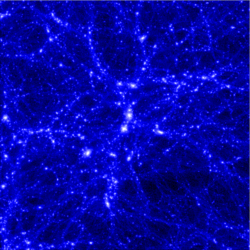}&
\includegraphics[scale=0.59]{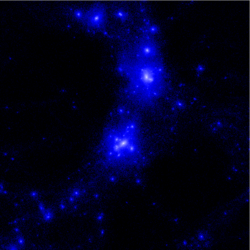} \\
\includegraphics[scale=0.59]{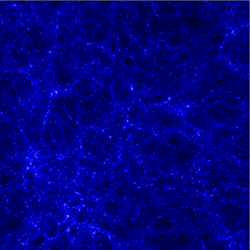} &
\includegraphics[scale=0.59]{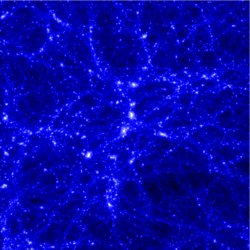}&
\includegraphics[scale=0.59]{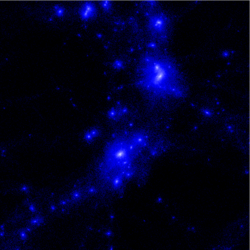} \\
\includegraphics[scale=0.59]{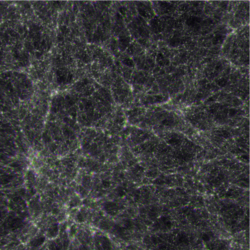} &
\includegraphics[scale=0.59]{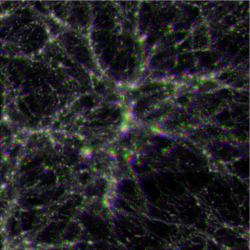}&
\includegraphics[scale=0.59]{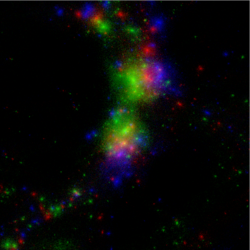} \\
\end{tabular}

\caption{
 Projections of the DM density fields for the $\Lambda \rm CDM$
  (first line), $\rm SUCDM$ (second line) and $\rm RPCDM$ (third line). 
  Left panels correspond to one full
  simulation box of $162\rm \; h^{-1}Mpc$ length projected on the x-y plane, central and right panels respectively     
  illustrate a cube of $32 \rm \;
  h^{-1}Mpc$ and $6 \rm \; h^{-1}Mpc$ lengths extracted from the full simulation ($162\rm \; h^{-1}Mpc$) and then
  projected on their x-y plane.  
  The color levels then indicate the surface density of these projections
  (in units of the cosmological mean surface density) with ranges $[2.03\times 10^{-1};6.41\times 10^{1}]$, $[2.63;8.32\times 10^{2}]$,
  $[6.33\times 10^{1};1.26\times 10^{4}]$ for the boxes of $162\rm \; h^{-1}Mpc$ (left), $32\rm \; h^{-1}Mpc$ (center), $6\rm \; h^{-1}Mpc$  (right) respectively.
  The last three panels in the bottom line show the superposition of the DM density field for
  the three different cosmologies with color encoding: green ($\Lambda
  \rm CDM$), red (SUCDM) and blue (RPCDM). On large scales, the differences
  are small (black and white regions), while on small scales these
  are clearly obvious (multi-color region).}
\label{snapshots}
\end{center}
\end{figure}

\subsection{Simulations set}

The scope of this series of articles is to accurately investigate the imprints of realistic dark energy
models on the cosmic structure formation. To this end we perform a set
of simulations with different box lengths, which allows us to precisely
evaluate the dark energy effects at different scales. For
numerical reasons, we need to comprimise between a spatial resolution
small enough to follow the formation of halos, and a volume box
sufficiently large such as to provide sufficient statistics. 
We have therefore run simulations for three different box sizes,
probing scales from kpc to Gpc: the largest box with $L=1296\rm
 \;h^{-1}Mpc$ provides us a good statistics on cluster counts; while
the smallest box with $L=162\rm \;h^{-1}Mpc$ has sufficient spatial
resolution for accurate measurements of the halo density profiles.
An intermediate box with $L=648\rm\;h^{-1}Mpc$ is a good compromise
for probing both the linear and non-linear scales of the matter power
spectrum which we discuss in this paper. Because of the spatial and
mass resolution we expect very small
finite volume effects (Takahashi et al, 2008). 
We want to remark that the results
presented in this paper our consistent with those derived using the 
$1296\rm \;h^{-1}Mpc$ and $162\rm \;h^{-1}Mpc$
simulations boxes, thus reinforcing our conclusions. 

The nine simulations that we have performed have a very large
 resolution compared to that of previous studies on structure
 formation in dark energy cosmologies. Each simulation contains $512^3$
particles, $512^3$ grids elements on the coarse level with $6-7$
additional refinement levels. Each simulation required about $20.000$ 
hours of computation on $64$ processors, while the more structured
simulations demanded $300$ hours of elapsed time.

Table~\ref{tab3} summarizes the characteristics of
the simulations for each of the three cosmological models.

\begin{table}
\begin{center}
\begin{tabular}{cccc}
\toprule
Parameters  & $162\rm\; h^{-1}Mpc$ & $648\rm\; h^{-1}Mpc$ & $1296\rm\; h^{-1}Mpc$\\
\midrule
$z_{\rm in }$ & $93$ & $56$ & $41$ \\
$m_p \rm (h^{-1}\;M_\odot)$ & $2.28\times 10^9$ & $1.46\times 10^{11}$ & $1.17\times 10^{12}$ \\
$\Delta_x \rm (h^{-1} kpc)$ & $2.47$ & $19.78$ & $39.55$ \\
$l_{\rm max}$ & $7$ & $6$ & $6$ \\
\midrule
$z_{\rm in }$ & $81$ & $50$ & $37$ \\
$m_p \rm (h^{-1}\;M_\odot)$ & $2.02\times 10^9$ & $1.30\times 10^{11}$ & $1.04\times 10^{12}$ \\
$\Delta_x \rm (h^{-1} kpc)$ & $2.47$ & $19.78$ & $39.55$ \\
$l_{\rm max}$ & $7$ & $6$ & $6$ \\
\midrule
$z_{\rm in }$ & $92$ & $55$ & $40$ \\
$m_p \rm (h^{-1}\;M_\odot)$ & $2.20\times 10^9$ & $1.41\times 10^{11}$ & $1.13\times 10^{12}$ \\
$\Delta_x \rm (h^{-1} kpc)$ & $2.47$ & $19.78$ & $39.55$ \\
$l_{\rm max}$ & $7$ & $6$ & $6$ \\
\bottomrule
\end{tabular}
\caption{Simulation parameters for the $\Lambda \rm CDM$, $\rm RPCDM$
  and $\rm SUCDM$  cosmologies from top to bottom. Each simulation contains $512^3$ particles
with $512^3$ coarse-grid cells and has been evolved up to $z=0$. The
  various entries report the values of the simulation initial redshift
  $z_{\rm in}$, particle mass $m_p$, spatial resolution $\Delta_x$ and
  maximal number of refinements $l_{\rm max}$.}
\label{tab3}
\end{center}
\end{table}

\subsection{Numerical results}

A visual summary of the N-body simulations is shown in 
Figure~\ref{snapshots}, where we plot for each cosmology the projected dark matter
density on the $x-y$ plane at $z=0$ from the simulations of box length
$162$~h$^{-1}$Mpc (panels from top to bottom correspond to $\Lambda$CDM, SUCDM and RPCDM respectively) 
at scales $162\rm \; h^{-1}Mpc$ (first column panels), $32\rm \;
h^{-1}Mpc$ (second column panels) and $6\rm \; h^{-1}Mpc$ (third
column panels). We may notice the high spatial resolution of these
simulations which probe both cosmological scales, filaments, halos and even the
inner profile of dark matter halos. 

In the images we have kept the same color intensity coding which facilitates a visual appreciation of the differences
of the clustering process for the different cosmologies. As already
found by F\"uzfa and Alimi (2006), realistic quintessence
cosmologies yield a DM density field that is less structured 
than that of the concordance $\Lambda$CDM, especially on small scales.
The visible features of the density field also show that DM structures 
have undergone a different evolution in terms of collapse 
and merging history. For instance this can be seen by comparing the DM
density distribution of the three models in the zoomed images at
$6\rm \; h^{-1}Mpc$. Such differences are emphasized in the bottom panels of
Figure~\ref{snapshots} where we plot in different colors 
(while keeping fixed the coding of the color intensity with the density field) the
superposition of the DM density field of the three cosmologies: green ($\Lambda$CDM), red
(SUCDM) and blue (RPCDM). The white regions correspond to an exact
surposition of the primary colors, thus indicating regions which have
undergone similar clustering. From the panels in the top raw
we can see that on the large scales the DM clustering
is nearly similar for the three different models, since it is
dominated by a white color. In contrast, when
zooming on smaller scales we can see the emergence of the individual
color associated with the density field of each model. In particular 
we may notice that the DM structures are
modified in shape, size and location.

\begin{figure}
\begin{center}
\includegraphics[scale=0.5,angle=0]{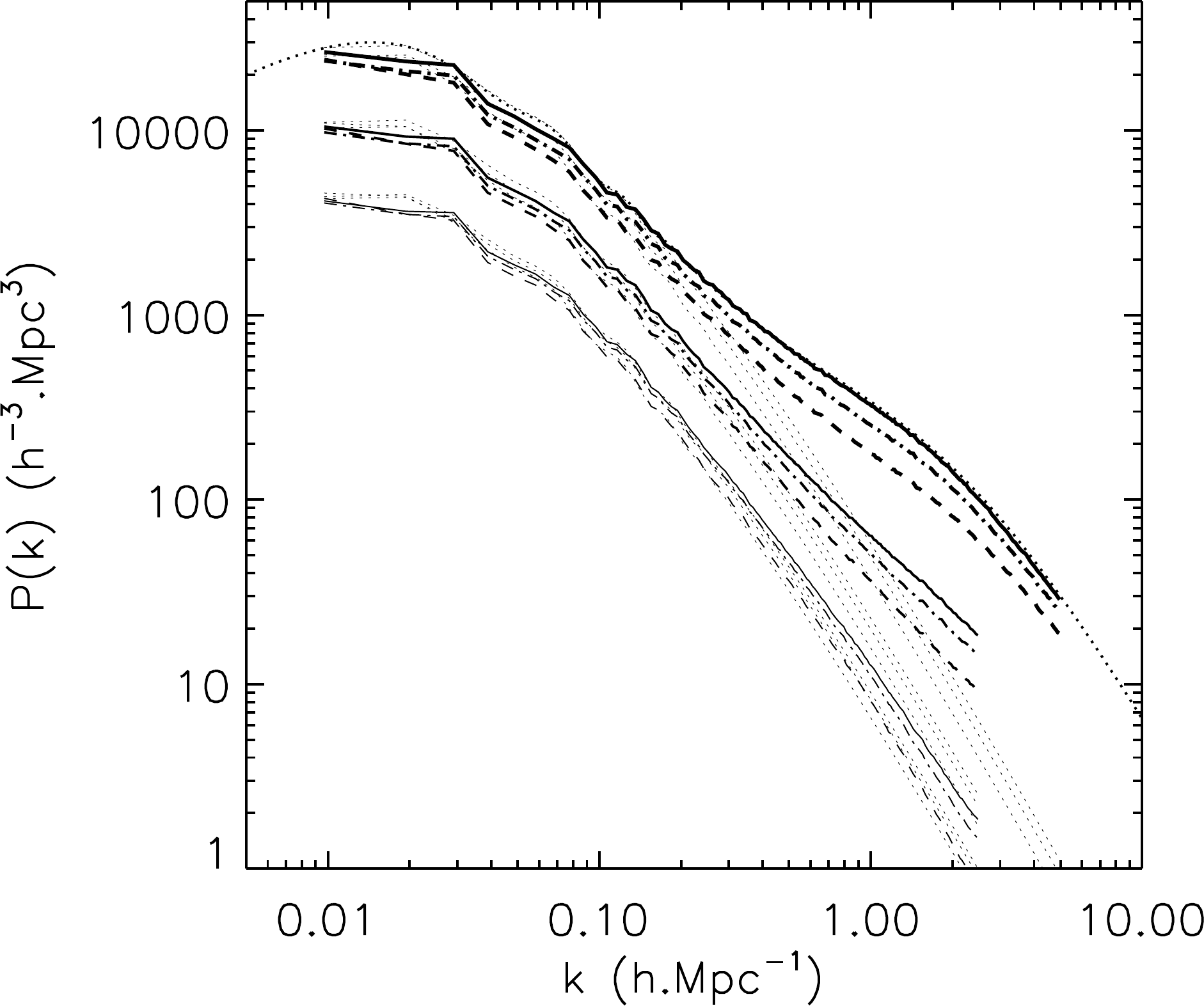}
\caption{Comparison between linear (dot lines) and non-linear power spectra
  from the N-body simulations with box length of 648~h$^{-1}$Mpc for the different cosmologies ($\Lambda
  \rm CDM$: solid line;$\rm RPCDM$: dash line; $\rm SUCDM$: dash-dot line) at three different epochs (from bottom to top: $a=0.3,\; 0.5,\; 1$).
  For comparison we plot the Smith et al. (2003) predictions of the power
spectrum at z=0 for the $\Lambda
  \rm CDM$ cosmology (top dotted line).
}
\label{pk_nonlin}
\end{center}
\end{figure}

\begin{figure}
\begin{center}
\includegraphics[scale=0.5,angle=0]{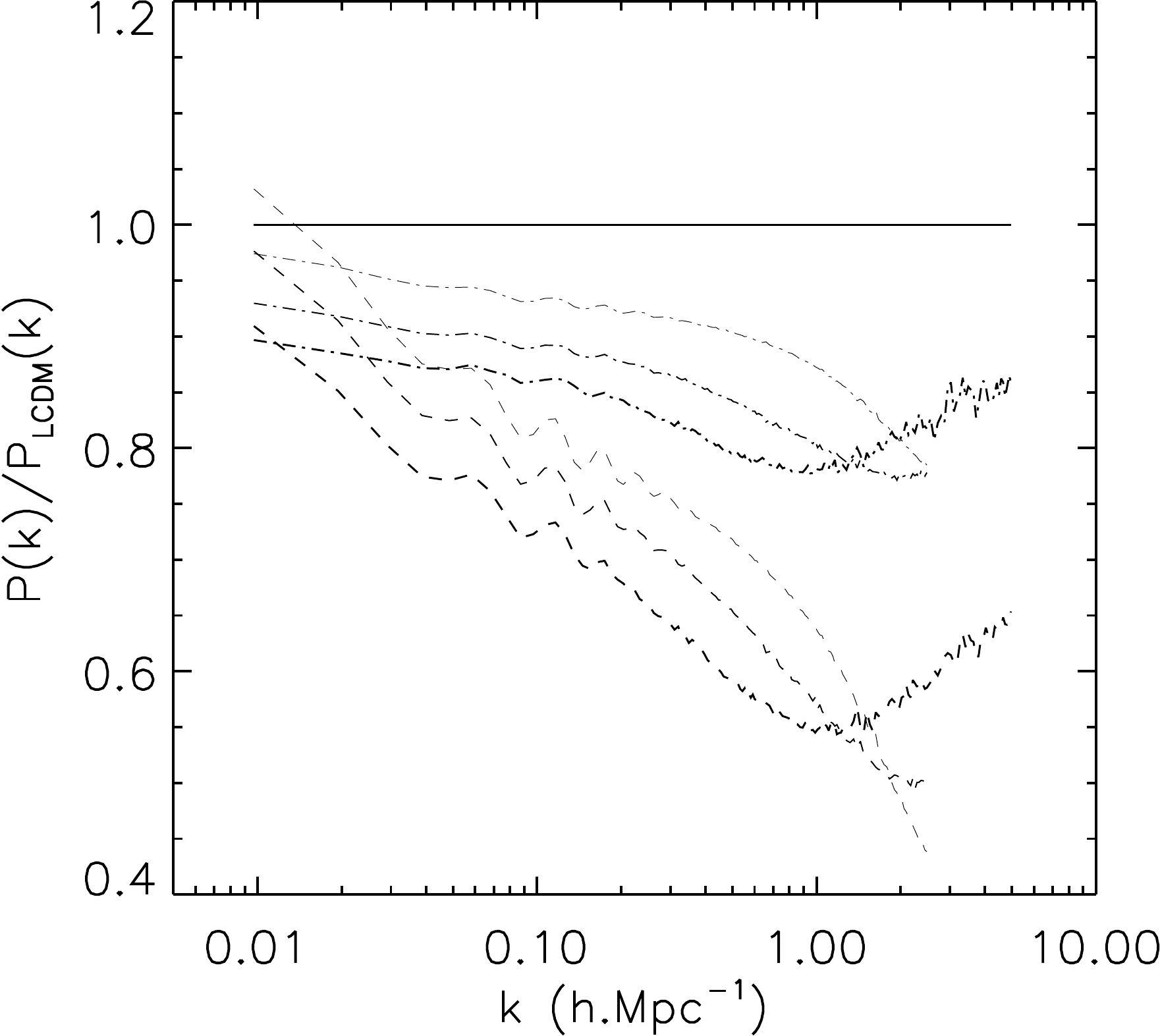}
\caption{Evolution of the non-linear matter power spectrum in quintessence cosmologies relative to the $\Lambda \rm CDM$ case
($\rm RPCDM$: dash line; $\rm SUCDM$: dash-dot line). From top to bottom: $a=0.3,\; 0.5,\; 1$, the thickness of the curves is proportional to the scale factor.
 }
 \label{pk_3}
\end{center}
\end{figure}

In order to quantify such differences for each simulation box we measure the matter power
spectrum using the \textsc{POWMES} code\footnote{We thank Stephane Colombi
  for kindly providing us with the code.} described in great
detail in Colombi et al. (2009). 
This code relies on a Taylor expansion of the
trigonometric functions and turns out to provide a very accurate estimation of
power spectra for N-body experiments. For all snapshots, we have used an
expansion of order 3, with a grid of size $256^3$, a number of foldings of 3
(to probe high wavenumbers) without substraction of the shot noise
(since in this paper we mainly focus on ratios of power spectra). For the
range of wavenumbers considered, the errors in the power spectrum estimation
are negligible compared to the approximations of the N-body solver.

The measured non-linear power spectra in the simulations of box length
$648$~h$^{-1}Mpc$ are plotted in Figure~\ref{pk_nonlin} for redshifts $z=0,1$ and
$2.3$ (i.e. $a=1,0.5$ and $0.3$). For comparison we also plot the
corresponding linear power spectra evolved backward using the homogeneous
growth function as described in Section~\ref{quinte_impr}. To be conservative we restrict
ourself to wavenumber below the Nyquist frequency, which is given by our maximum resolution at a
given redshift divided by a chosen value of 32. In this way, our highest wavenumber is of order
of one or two times the Nyquist frequency associated to the coarse
grid. According to the cosmic code comparison analysis by Heitmann et
al. (2005, 2008), the majority of the tested N-body codes agree at
$~5\%$ level on scales which are below twice the Nyquist frequency of the coarse
grid. However, above four times this frequency the agreement becomes worse and deviations
can be larger than ten percent, this is why we choose the more conservative criterion
described above. Moreover, we focus on the power spectrum below $k=5$~h Mpc$^{-1}$
since at $z=0$ the baryon contribution is of a few percents at this
scale and increases up to 10 percents at $k=10$~h Mpc$^{-1}$ (Jing et al,
2006). We note that these uncertainties only concern the absolute prediction
of the power spectrum and not the ratio of power spectrum which we will describe in Section 5.5. 

 Overall quintessence models possess less
power than $\Lambda$CDM, which confirms the expectation of the linear
analysis. Again this can be interpreted as a consequence of the
fact that a record of dark energy during the linear collapse is kept
through the non-linear regime. We may notice 
several important features. Firstly, the amplitude of the bumps on the 
large scales grows earlier in $\Lambda$CDM, since
the large scales in the simulation boxes evolve nearly linearly, then this 
result is consistent with the model differences expected from the
linear growth factor discussed in Section~\ref{quinte_impr}. 
Secondly, we can see that at small scales 
the departure from linearity occurs earlier in $\Lambda$CDM than 
in quintessence models. Again this is consistent with the fact that during 
the matter dominated era the concordance model decelerates more 
efficiently than quintessence cosmologies (see also bottom panel of Figure \ref{wq}), causing a faster rate of gravitational
infall of dark matter, thus entering in the non-linear regime at
earlier times. 

One last point concerns the detectability of such differences through
measurements of the galaxy power spectrum. 
On the large scales the measured power spectra matches the
linear predictions which we have shown to be in agreement with SDSS
data provided the large scale galaxy bias parameter assumes slightly
different values for each realistic cosmology. In order to detect the
imprint of dark energy on the non-linear scale of the galaxy power
spectrum is therefore necessary to have an accurate knowledge of the
galaxy bias on those scales. In such a case measurements of the weak
lensing power spectrum probing the gravitating mass distribution can
be more reliable (see e.g. Huterer 2002). In the upcoming years
several weak lensing surveys will provide accurate tomographic
measurements of the shear power spectrum which can be especially
sensitive to the dark energy imprint on the small scales distribution
of dark matter halos (Ivezic et al. 2008).

\subsection{Imprints of dark energy on cosmic structure formation}

In this section, we aim to provide a detailed physical understanding of
the imprint of the three realistic models on the non-linear power
spectrum. For this reason rather than focusing on the power spectrum
of each model we focus on their ratios. This has also an advantage
from a purely numerical point of view, since McDonald et al. (2006)
have investigate the sensitivity of power spectra
to changes of the starting redshift, box size, mass resolution, force
resolution and time step size and shown that most of the errors cancel out
when considering ratios of power spectra, with an expected precision of a few percents around
the Nyquist frequency of the coarse grid.

In Figure~\ref{pk_3} we plot the ratio of the measured non-linear power
spectrum of the RPCDM (dash line) and SUCDM (dash-dot line) 
to that of the concordance $\Lambda$CDM, $r=P_{nl}^{\rm QCDM}(k)/P_{nl}^{\Lambda\rm CDM}(k)$. 
All quintessence cosmologies are characterized by $r<1$, corresponding
to a smaller amount of clustering with respect to $\Lambda$CDM.
We may notice that the discrepancy with respect to the concordance model is not uniform, 
since it varies with scale, redshift and quintessence model. Overall the RPCDM differs from the $\Lambda$CDM more than
the SUCDM, with a maximum discrepancy of about $40\%$ for the RPCDM
and about $20\%$ for the SUCDM. Besides in both quintessence
cosmologies $r$ decreases as function of $k$ in the interval
$0.01 \; \rm h \;Mpc^{-1}\lesssim k\lesssim 1 \; \rm h \;Mpc^{-1}$,
and increases at smaller scales, $k> 1 \; \rm h \;Mpc^{-1}$.

This is consistent with the fact that on the large scales the linear growth rate
is suppressed with respect to the $\Lambda$CDM case more in the RPCDM than SUCDM (see Fig.~\ref{sigma}).
On the other hand the small scales are affected by non-linear
processes which increases the amount of clustering, thus inverting the trend. 
We can also notice that as consequence of the non-linear collapse at a given time there exists a
characteristic non-linear scale (corresponding
to the minimum of the $r$-curves) below which the DM clustering increases
as function of $k$. Such a scale shifts from the 
right to the left (from $k\approx 1\; \rm h \;Mpc^{-1}$ to $\approx 3\; \rm h \;Mpc^{-1}$) 
for increasing redshift, i.e. the smaller the scale the earlier
it enters in the non-linear regime. Furthermore, the amount of clustering on the non-linear scales 
in the RPCDM model differs from the SUCDM case, thus indicating a chracteristic imprint of the nature of DE.

\begin{figure}
\begin{center}
\begin{tabular}{c}
\includegraphics[scale=0.4,angle=0]{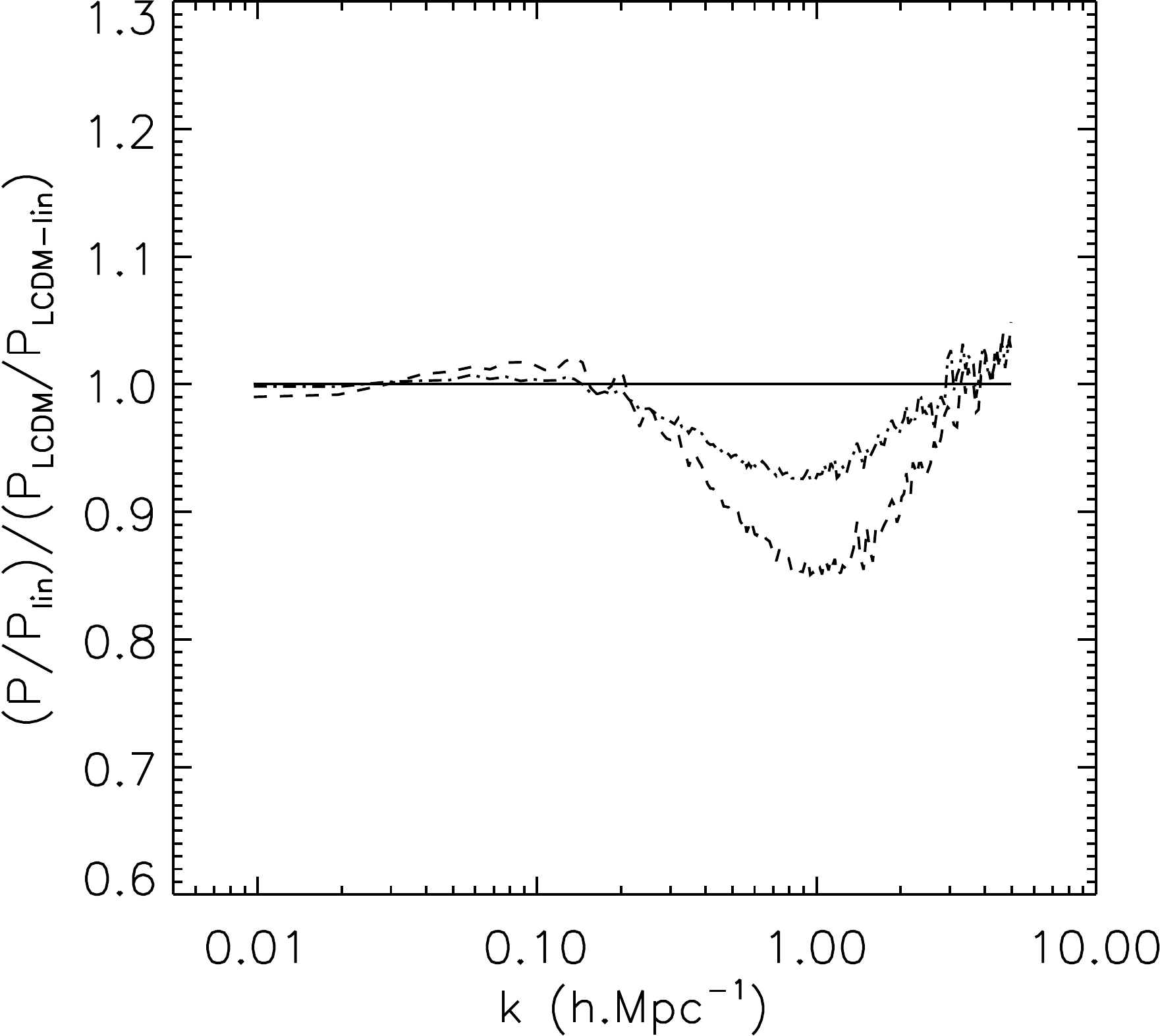}\\
\includegraphics[scale=0.4,angle=0]{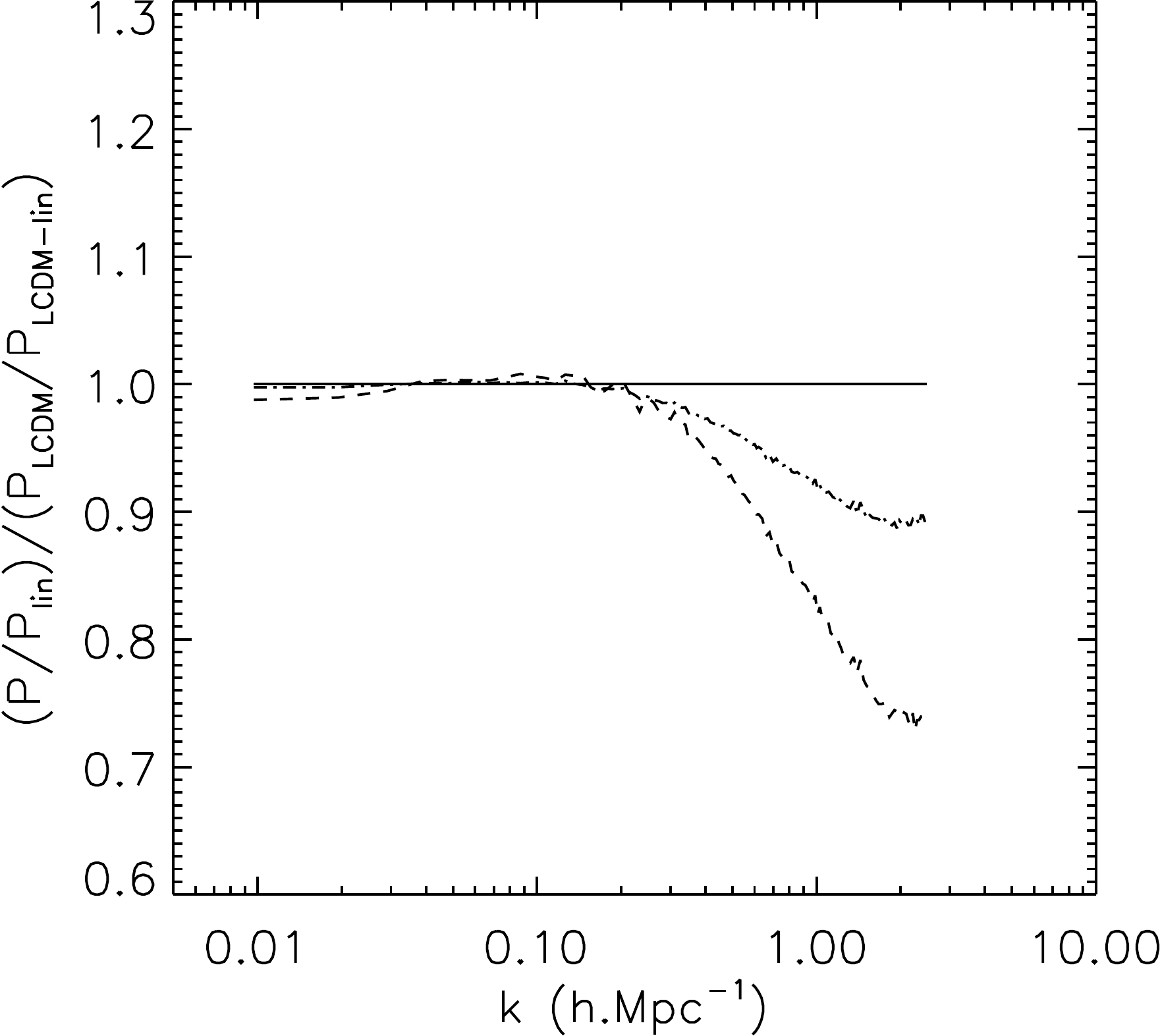}\\
\includegraphics[scale=0.4,angle=0]{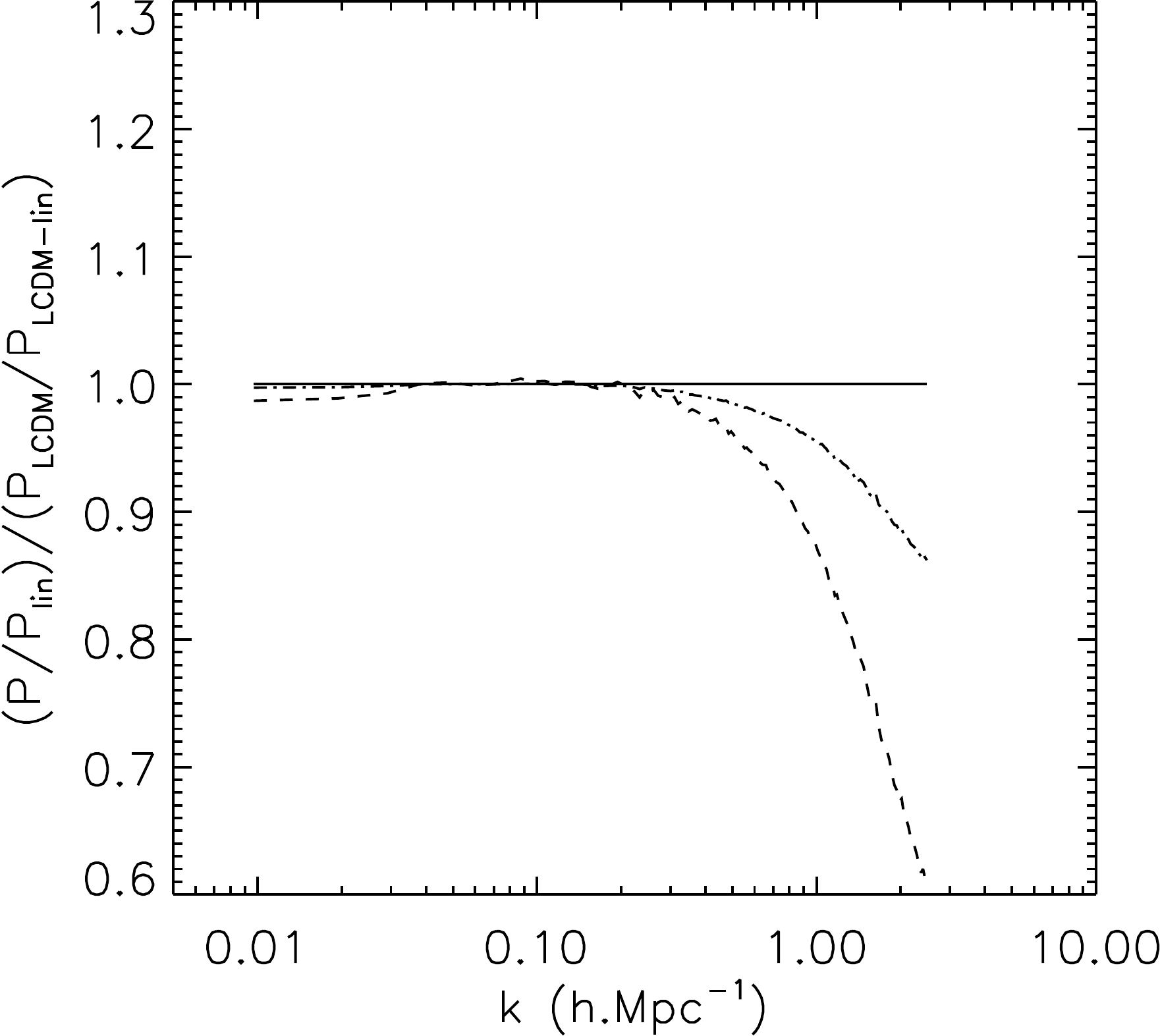}
\end{tabular}
\caption{Ratio of the non-linear power spectrum relative to linear predictions for the different cosmologies as a measurement of the evolution of non-linearity
in the gravitational collapse. Panels from top to bottom corresponds
to $a=1,\; 0.5,\; 0.3$ ($\Lambda \rm CDM$: straight line;
$\rm RPCDM$: dashed line; $\rm SUCDM$: dash-dotted line). 
 }
 \label{pk_2}
\end{center}
\end{figure}

In order to isolate the non-linear contributions and remove the linear effects we compute the ratio:
\begin{equation}
R_{\rm QCDM}=\frac{P_{nl}^{\rm QCDM}(k)}{P^{\Lambda\rm CDM}_{nl}(k)}\;\frac{ P_{lin}^{\Lambda \rm CDM}(k)}{P^{\rm QCDM}_{lin}(k)},
\end{equation}
which is shown in Figure~\ref{pk_2} for $a=1,0.5$ and $0.3$ respectively. 
We can see that all scales corresponding to $k\lesssim 0.1-0.2\;\rm h \;Mpc^{-1}$ are
in the linear regime, since $R_{\rm QCDM}\approx 1$ (within $1-2\%$
accuracy). On smaller scales corresponding to $0.1\;\rm h \;Mpc^{-1} \lesssim
k\lesssim 1\;\rm h \;Mpc^{-1}$, the ratio decreases for both quintessence
models with respect to the $\Lambda$CDM case, as already seen in Figure~\ref{pk_3}. This suggests that also non-linear effects
contribute to the power suppression that characterizes quintessence
models. This corresponds to the quasi-linear regime. On the
other hand on scales which are below the scame corresponding to the location of the minimum of the
$R_{\rm QCDM}$ function, $k\approx 1\;\rm{ h} \;Mpc^{-1}$, 
the non-linear clustering of DM in quintessence models increases very rapidly, eventually
exceeding that of the $\Lambda$CDM ($R_{\rm QCDM}>1$). In other words as time elapses the non-linear growth
of structure becomes more and more efficient in QCDM models than
in $\Lambda$CDM, with the RPCDM having a larger growth than the
SUCDM. This can be understood in terms of the stable
clustering regime characterizing these scales (Hamilton et al. 1991) or alternatively to the halo term
(Smith et al. 2003), such that the growth of the power spectrum tends to slow down once the structures become more and more virialised.

\begin{figure}
\begin{center}
\begin{tabular}{c}
\includegraphics[scale=0.4,angle=0]{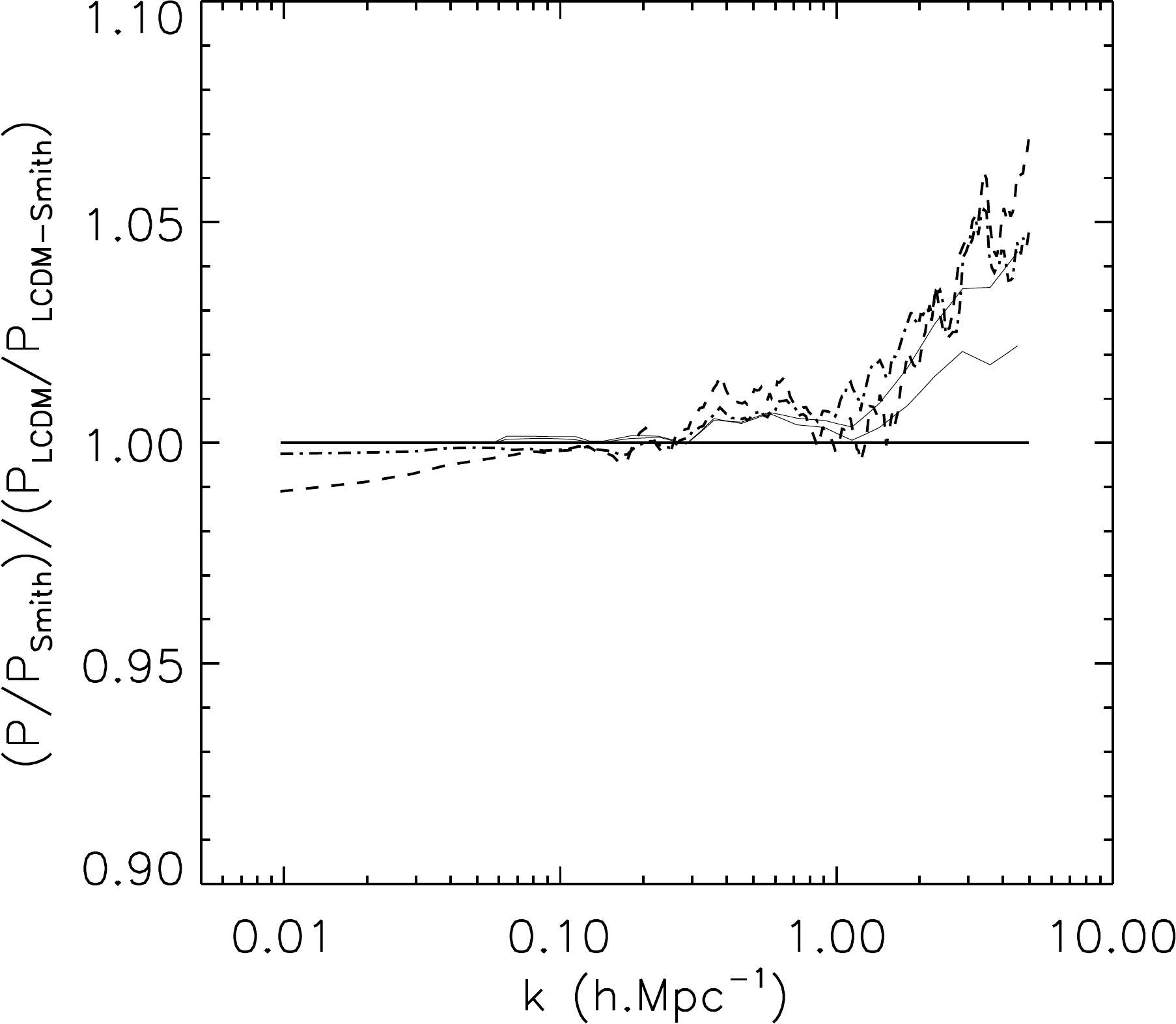}\\
\includegraphics[scale=0.4,angle=0]{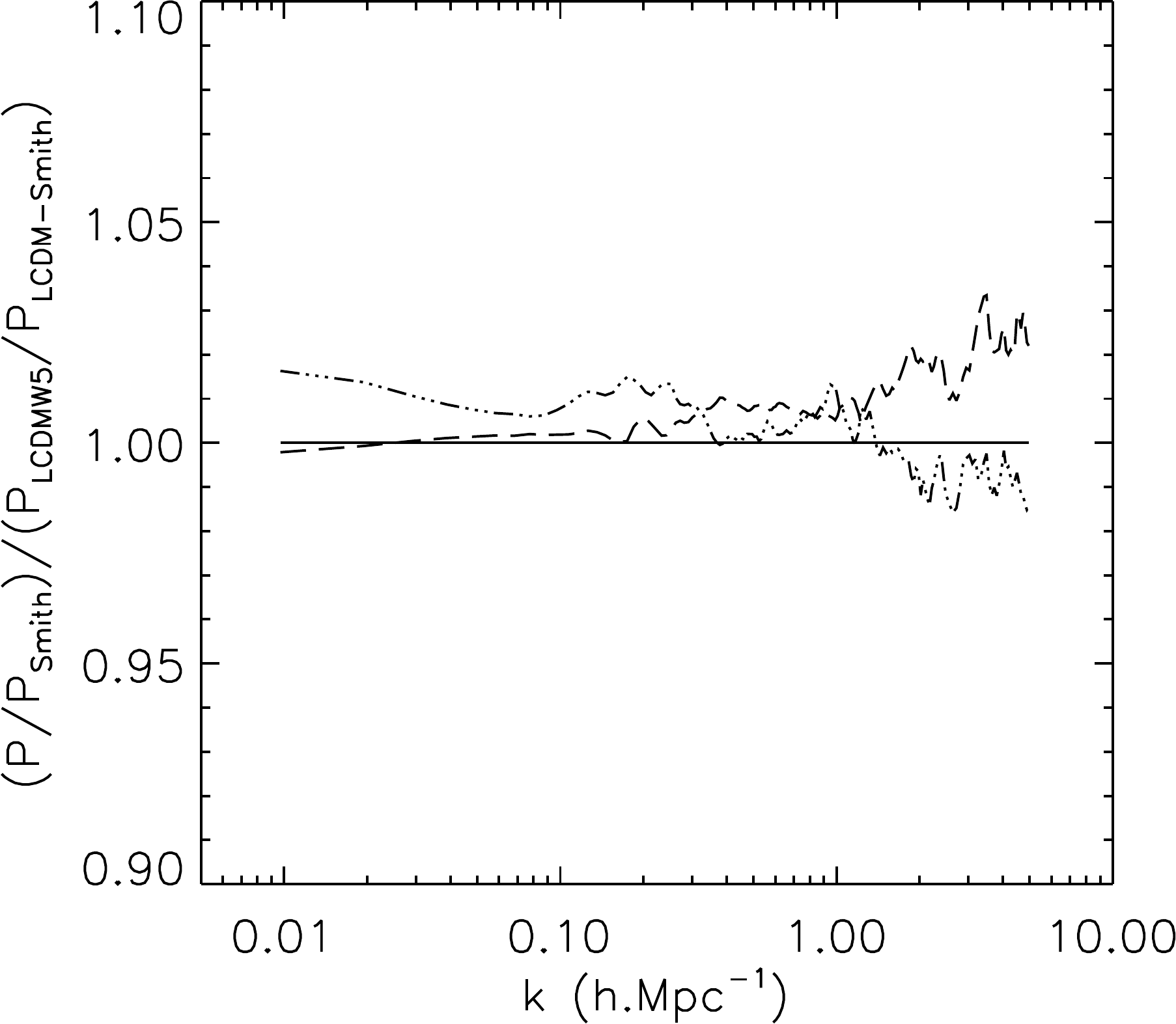}\\
\includegraphics[scale=0.4,angle=0]{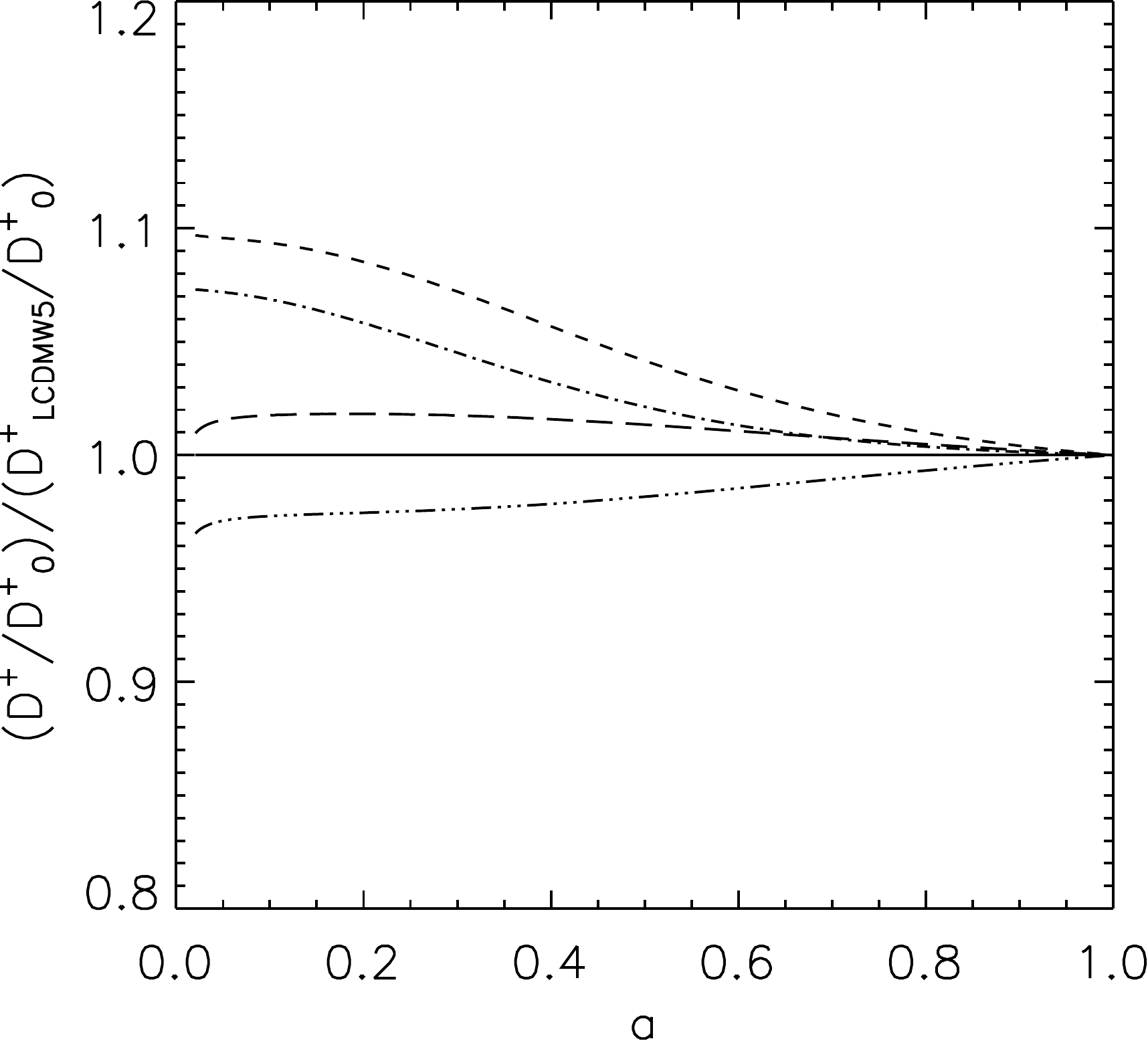}
\end{tabular}
\caption{Top panel: Ratio of the non-linear power spectrum normalized
  to  Smith et al. (2003) fit for RPCDM (dashed) and SUCDM
  (dot-dashed) relative to the $\Lambda$CDM. The thin continuous line are the constant eos
  expectation for RPCDM (top) and SUCDM (bottom). Middle pane: as in
  the top panel but for the $\Lambda$CDM WMAP I (triple dot-dashed line) and $\Lambda$CDM WMAP III (long
  dashed line) models. Bottom panel: Linear growth factor relative to the $\Lambda$CDM WMAP5
  one. The deviations at high k of the power spectra are correlated with the linear growth histories.}
 \label{history}
\end{center}
\end{figure}

At this point, we have isolated three main contributions to the rough differences between the
power spectra of different dynamical dark energy models. On linear
scales ($k<0.1$ h Mpc$^{-1}$) 
the discrepancies can be explained by the
differences in the linear power spectra (which are due to the various
cosmological parameters needed to reproduce the CMB and SNIa, the spatial
fluctuations of the quintessence as well as the different background
evolutions). On quasi-linear scales ($0.1$ h Mpc$^{-1}< k <1$ h Mpc$^{-1}$),  the discrepancies can be mainly explained by the
non-linear amplification of the growth rate. Since the linear amplitude
is different, a given mode enters the non-linear regime at a different
time. On the halo scales ($k>1$ h Mpc$^{-1}$) , the
differences are roughly due to the saturation of the growth of structures as they
start to be virialised. Again, this halo term depends mainly on the amplitude
of the linear growth rate. 
\\
\\
All the effects described above are rather well accounted for by the
Smith et al. (2003) fitting function of the matter power spectrum.
In order to check whether there are additional non-linear contributions to the final
shape of the power spectrum, we compute the ratio
\begin{equation}
RS_{\rm QCDM}=\frac{P_{nl}^{\rm QCDM}(k)}{P^{\Lambda\rm CDM}_{nl}(k)}\;\frac{ P_{Smith}^{\Lambda \rm CDM}(k)}{P^{\rm QCDM}_{Smith}(k)},
\end{equation}
plotted in Figure~\ref{history} (top panel) for $z=0$. We can see that on scales
$k>1 \rm h\; Mpc^{-1}$ (halo scale) there is an enhancement of the power in quintessence
models compared to the Smith et al. (2003) prediction. This can be as
large as $5\%$ at $k=5 \rm h\; Mpc^{-1}$ and reach even
larger values at higher wavenumbers. This deviation is not a numerical
artifact because as we already mentionned above, these wavenumbers are below the Nyquist
frequency. Moreover the ratio of
power spectra between two models is less sensitive to numerical errors that
the absolute power spectrum itself, as previously mentioned.

This result is qualitatively consistent with a number 
of works (McDonald et al. 2006; Ma 2007; Francis et al. 2007 and Casarini et al.
2009). For instance, in Figure~\ref{history} we plot the deviations for SUCDM (dash-dotted line)
and RPCDM models and for comparison those tabulated by McDonald et al. (2006) (thin
continuous lines) in the case of dark energy models parametrized by
constant equation of state $w=w_0$ (see table \ref{tab2} for the
values of $w_0$). We find that deviations are nearly of the same order, $~5\%$ level, towards higher values of the
power spectrum at large k. Nonetheless the are some important differences. First of all, the deviations
appear to be stronger in dynamical quintessence models than in the
$w=\rm const.$ case. Secondly, the shape is slightly
different near $k=5 \rm h Mpc^{-1}$. Finally, the disagreement is
stronger for the SUCDM cosmology. The reason is rather obvious if we
look at Fig.~\ref{wq} (top panel). In fact the time variation of the equation of state is larger in
SUCDM. Recently, a more sophisticated way of fitting these deviations has
been proposed for a linear parametrization of the equation of state (Francis et al. 2007) and
extended to quintessence models (Casarini et al. 2009). The idea is to use
the prediction of McDonald et al. (2006) for an effective constant eos $w_{eff}$,
that predicts the same distance to the last scattering surface of a dynamical quintessence model. 
Although interesting, the generality of this approach is not clear. 
In addition, the mapping between the quintessence model and the effective description with constant eos
at different epochs is not straighforward to implement, and can easily lead to misinterpretations. 
In comparison, the complete approach described here is physically well
motivated and allows for a robust accounting of the various
effects. As we have shown here, each dark energy model has a specific
cosmological evolution which leads to specific imprints, and it is not clear
how well these features can be precisely accounted for in the approach
proposed by Casarini et al. (2009).

This point is also related to another important aspect, namely the universality
of fitting procedures of the non-linear power spectrum based on large numerical
simulations. As already pointed out by Ma (2007), the history of structure formation is not
properly taken into account by standard predictions such as that
provided by Smith et al. (2003) or Peacock \& Dodds (1996). This is because the non-linear power
spectrum is expressed as a fitting function of the cosmological parameters
and the linear power spectrum. To some extent these fitting functions
can be considered as instantaneous predictions from the linear
power spectrum, since they do not depend in any way on the past
evolutionary history of the structure formation of the model specified by the
set of cosmological parameters. However, the non-linear regime should
in principle have a fossil record of such history and thus the past
influence of dark energy. As an example Ma (2007) 
has shown that the deviations from one model to another scale as their
relative linear growth histories. Indeed, the imprint of the past
evolution on the non-linear power spectrum is also present in the case
of the realistic models considered here. In order to emphasize this
result we have run two additional $\Lambda$CDM simulations for comparison: LCDMW1 (WMAPI
cosmological parameters: $\Omega_m=0.29$, $\Omega_{\Lambda}=0.71$,  $\Omega_b=0.047$, $h=0.72$,
$\sigma_8=0.9$ and $n_S=0.99$ ) and LCDMW3 (WMAPIII cosmological parameters:
$\Omega_m=0.24$, $\Omega_{\Lambda}=0.76$,  $\Omega_b=0.042$, $h=0.73$,
$\sigma_8=0.74$ and $n_S=0.951$).
The deviations with the respect to the Smith et al. (2003) prediction
are shown in Figure~\ref{history} (middle panel). We may note that the
curves are in remarkably good agreement, although some small deviations
(below $5\%$) are present. The different level of deviations between the realistic
quintessence models and these $\Lambda \rm CDM$ test runs can be
explained in terms of their past history as described by the linear
growth factor evolution. In Figure~\ref{history} (bottom panel) we
plot the linear growth factor as function of the scale factor relative
to the $\Lambda \rm CDM$ WMAP5 case. We can see that deviations 
with respect to Smith et al. (2003) at high k are well correlated to the 
maximum deviations between the linear growth histories of the
different models. 
Thus indicating that the past history of cosmic structure formation plays an important role since the features introduced in the matter power spectrum are not erased by the non-linear collapse: the non-linear collapse contributes to the specific imprints of the DE model.
 A semi-analytical implementation of the
structure formation history on the calculation of the non-linear power
spectrum can be obtained using the halo model (Cooray \& Sheth
2002), which accounts for the instant of collapse, the overdensity at virialization and the concentration parameters
which depends on the assembly history (Wechsler et al. 2002). This is further
motivation for us to study in great details the impact of dark energy on the
mass function and on the internal structure of dark matter halos,
which will be discussed in other articles of the series.

\section{Conclusions}\label{conclu}
In this paper, we have studied the imprint of quintessence on the
non-linear clustering of dark matter halos through state-of-the-art N-body
simulations. To this purpose we have focused on two scalar field models specified
by Ratra-Peebles and SUGRA potentials. Upon these models we have
performed a series of high resolution N-body simulations and shown
that quintessence leaves a distinctive signature on the structure
formation at all scales and especially during the non-linear regime.
This is in contrast with conclusions drawn from previous works in the
literature. Some of these studies limited their analysis to dark
energy models parametrized by a simply constant or linearly varying
equation of state, which fail to grasp the specific feature of dark
energy clustering throughout the cosmological evolution. Other
analysis have focused on the same class of 
quintessence models; nevertheless, in all these studies 
the cosmological parameters were set to values
corresponding to the concordance $\Lambda$CDM model, thus missing
crucial features which are necessary to fully evaluate the impact of dark energy
on the structure formation. 

The benchmark of dark energy models considered has been selected through
a likelihood data analysis which we performed to determine
quintessence model parameters that fit CMB and SN Ia data within
$2\sigma$ significance from the $\Lambda$CDM scenario. 
These models are statistically indistinguishable from the concordance
model; nevertheless they are characterized by different cosmic
expansion histories, which in addition to the effect
of dark energy perturbations modify the growth of the linear dark
matter perturbations, thus leaving quintessence model dependent
signatures on the linear matter power spectrum.
These imprints are consistent with current measurements of the galaxy
power spectrum from SDSS data, and statistically 
indistinguishable with the respect to the
$\Lambda$CDM model.

The results of the N-body simulations clearly show that the model
dependent features present at the homogeneous and linear perturbation level
are amplified by the non-linear collapse of dark matter structures.
We find that quintessence affects the non-linear matter power 
spectrum at all scales, with a suppression of power at the small
scales compared to the $\Lambda$CDM case. However when 
the effects of the different linear evolution histories are taken into account,
the non-linear clustering of DM appears to be more efficient
in quintessence cosmologies than in the standard $\Lambda$CDM scenario.
These features can only be partly understood in the Smith et
al. (2003) phenomenology framework, that as we have shown here it
needs to be complemented with the full cosmic structure formation history
to account for the imprint of DE on the non-linear matter power
spectrum. This support the idea that a robust evaluation of the impact
of DE on the non-linear structure formation requires an accurate and
self-consistent implementation of the quintessence model in the N-body
simulations.

The imprints of quintessence on the DM distribution that we obtained here would affect the gas on
all scales. As the gas collapses inside DM potential wells, the gas distribution would also keep a record of the nature of DE.
On large scales, the gas distribution tracks the DM one while on small scales the gas cooling accelerates the gas collapse
but not the DM one. As a result, the star formation rate (SFR) could be affected by DE. From the results of this paper, the SFR  is
expected to be smaller in quintessential universes than in the $\Lambda\rm CDM$ model due to the smaller $\sigma_8$ in the first. However,
the SFR in the past could be more important compared to the SFR today when DE is quintessence rather than $\Lambda$. This is because the ratio
$D_+/D_+(a_0)$ in quintessence models is higher than in $\Lambda\rm CDM$. As well, the SFR at $z=0$
in quintessence models should be different than the one of a $\Lambda\rm CDM$ model with the same value of $\sigma_8$.
The history of SFR is therefore expected to keep a specific record of the cosmic expansion history and therefore of the nature of DE. 
More precisely, it has been shown by Rasera \& Teyssier (2006) that the total SFR in the Universe is ruled at first order by the amount
of DM halos with total mass higher than a critical Jeans-like mass called the filtering mass. This prescription could allow one
to deduce SFR from DM halo mass function (see the sequel of this paper). In addition, another expected imprint of DE on the gas would arise from the gas collapse itself whose details, like accretion, cooling, SFR, supernovae feedback, ..., depend on the cosmic expansion history. Finally, the feedback of the gas on DM will slightly affect DM distribution on small scales (see also Guillet et al., 2009).

In this work, we focused on the DE signature on the non-linear matter
power spectrum, in a series of upcoming papers we will present other characteristic
imprints of dark energy scenarios on the non-linear structure
formation, in particular we will provide 
a detailed study of the halo mass functions and 
the physical properties of dark matter halos.
The results of our analysis already suggest that, 
contrary to common belief, an accurate comparison between 
the small and large scale distribution of matter in the Universe 
can in principle shed light on the nature of dark energy.

\section*{Acknowledgements}
Numerical simulations for parameter selection were made at the UCLouvain HPC Center (Belgium) under project FRFC 2.4502.05, and on the local computing resources
at LUTh (Observatoire de Paris, France) and Unit\'e de Syst\`emes Dynamiques (FUNDP, Belgium).
% The N-body simulations presented here were performed at the  "Institut du D\'eveloppement et des Ressources en Informatique Scientifique" (IDRIS, France) and "Centre de Calcul
% Recherche et Technologie" (CCRT, CEA, France). 
This work was granted access to the HPC resources of CCRT under the
allocation 2009-t20080412191 made by GENCI (Grand Equipement National de Calcul
Intensif).
We would like to thank
Romain Teyssier, St\'ephane Colombi and Simon Prunet for their valuable advices concerning
the \textsc{RAMSES}, \textsc{POWMES} and \textsc{MPGRAFIC} softwares. This work was supported by the
Horizon Project (www.projet-horizon.fr).
A.F. is supported by the Belgian "Fonds de la Recherche Scientifique" (F.N.R.S. Postdoctoral Researcher) and is associated researcher at CP3 (UCL, Belgium)
and LUTh, Observatory of Paris. V.B. is supported by the Belgian Federal Office for Scientific, Technical and Cultural Affairs through the Interuniversity Attraction Pole P6/11.

\end{document}